\documentclass{article}

\usepackage{amssymb,bbm}
\usepackage{amsthm}
\usepackage{latexsym}
\usepackage{amsmath}
\usepackage{mathtools}
\usepackage{hyperref}
\usepackage{pdflscape}
\usepackage{booktabs}
\usepackage{multirow}
\usepackage{subfigure}
\usepackage{float}
\usepackage[T1]{fontenc}
\usepackage{xcolor}

\usepackage[ruled, linesnumbered]{algorithm2e}

\SetCommentSty{mycommfont}
\usepackage[backend=biber,natbib=true,bibencoding=utf8,style=authoryear]{biblatex}
\newcommand{\IP}{\mathbb P}

\newcommand{\IE}{\mathbb E}
\newcommand{\IR}{\mathbb R}

\newcommand{\IS}{\mathbb S}

\renewcommand\leq{\leqslant}

\renewcommand\geq{\geqslant}

\newcommand{\cP}{{\mathcal P}}

\newcommand{\indic}[1]{\mathbbm{1}_{#1}}
\renewcommand{\epsilon}{\varepsilon}
\renewcommand{\phi}{\varphi}

\SetKw{Continue}{continue}

\DeclareMathOperator*{\argmax}{arg\,max}
\DeclareMathOperator*{\argmin}{arg\,min}

\newcommand*\diff{\mathop{}\!\mathrm{d}}

\newtheorem{thm}{Theorem}[section]

\newtheorem{lemma}[thm]{Lemma}
\newtheorem{problem}[thm]{Problem}

\title{Joint Calibration of Local Volatility Models with Stochastic Interest Rates using Semimartingale Optimal Transport}
\author{Benjamin Joseph$^1$\thanks{This research has been supported by BNP Paribas Global Markets and the EPSRC Centre for Doctoral Training in Mathematics of Random Systems: Analysis, Modelling and Simulation (EP/S023925/1).}, Gr\'{e}goire Loeper$^2$, and Jan Ob\l{}\'{o}j$^3$ 
}
\date{%
$^1$ Mathematical Institute and Christ Church, University of Oxford\\ \texttt{benjamin.joseph@maths.ox.ac.uk}\smallskip\\
$^2$ BNP Paribas Global Markets\\ \texttt{gregoire.loeper@bnpparibas.com}\smallskip\\
$^3$ Mathematical Institute and St John's College, University of Oxford\\ \texttt{jan.obloj@maths.ox.ac.uk}
}

\begin{document}
\maketitle
\begin{abstract}
We develop and implement a non-parametric method for joint exact calibration of a local volatility model and a correlated stochastic short rate model using semimartingale optimal transport. The method relies on the duality results established in \cite{JLO23calibration} and jointly calibrates the whole equity-rate dynamics. 
It uses an iterative approach which starts with a parametric model and tries to stay close to it, until a perfect calibration is obtained. We demonstrate the performance of our approach on market data using European SPX options and European cap interest rate options. Finally, we compare the joint calibration approach with the sequential calibration, in which the short rate model is calibrated first and frozen. 
\end{abstract}
\section{Introduction}
In financial engineering, calibration refers to the process of adjusting the model's prices to the market prices. It is an essential task for pricing and hedging derivatives, often performed on an intraday basis. 

For this purpose, two class of models emerge: the parametric, and the non-parametric ones. A typical example of parametric model is the Heston model.
This type of model  allows simplified (semi-analytic) pricing, as well as a clear financial interpretation for each of its parameters; however, its calibration capabilities are limited and it may even struggle to calibrate to a number of instruments comparable to the number of parameters.
On the other hand, a non parametric model such as the local volatility model allows to match exactly a whole surface of European option prices, at the expense of requiring numerical simulations, and of lack of interpretability. Nevertheless, it is widely used across the industry.

Recently, a novel non-parametric approach to calibration exploiting semimartingale optimal transport (SOT) has been proposed, and applied for local volatility calibration in \cite{guo2019local}.  It builds on the methodology introduced in  \cite{tan2013optimal} and can also be seen as the stochastic extension of the celebrated time-continuous formulation of the classical OT problem by \cite{benamou2000computational}. 
This approach is very versatile, and was then extended to local-stochastic volatility models in \cite{guo2022calibration}, to general path dependent products and constraints in \cite{guo2021path},  and to the joint calibration problem of SPX and VIX options in \cite{guo2022joint}. The last paper in particular showcased the ability of this method to obtain an exact calibration which was known to elude standard parametric models (see \cite{guo2022optimal} for a general discussion on optimal transport based calibration).
We mention here an earlier work \cite{avellaneda1997calibrating}, where the authors followed a similar path to calibration, based on a combination of  non parametric exact calibration and regularization via entropic penalization.

All financial products exhibit a dependency to interest rates, as they involve future payments, and it is therefore natural to take into account the stochastic dynamics of interest rates in the calibration procedure.
OT based calibration has recently been adapted by the authors of the present study to the context of equity products with stochastic interest rates \cite{JLO23calibration}.
Different approaches can be used in this task, 
either opting for a {\it sequential calibration}: the interest rates are calibrated (or given) first, and then a hybrid equity/rates model is calibrated to products exhibiting a dependency to both rates and equity dynamics, or, going for a {\it joint calibration} where the full covariance process is calibrated at once using rates and equity market prices. 

In \cite{JLO23calibration} a general duality result has been given, that covers both approaches. An application to the former case of sequential calibration was then proposed. This problem is naturally less involved numerically than the joint calibration problem.

In this paper, we use the previous duality result to address the joint calibration problem, given prices of European call options on equity and caps on the short rate. The resulting model is driven by a pair of correlated Brownian motions and has for state variables the equity underlying and the short rate. The algorithm is tested against real market data. We also compare the performance of joint calibration against sequential calibration considered in \cite{JLO23calibration}.
\section{Formulation of Semimartingale Optimal Transport Calibration Problem}
Our method offers exact joint calibration of the stock price's process and the stochastic short rate processes to market prices of, respectively, European options, European calls and caps. We perform a non-parametric calibration of the drift and diffusion coefficients by recasting the calibration problem as a Semimartingale Optimal Transport (SOT) problem with discrete constraints (given by observed market data). Our SOT reformulation can be seen as a projection on to the space of exactly calibrated models: we start with a given reference model and the cost functional which we seek to penalize deviations from that reference. While enjoying the accuracy of exactly calibrating to chosen product prices, the penalization away from the reference model offers several advantages. First, it convexifies the problem giving a unique solution (for a fixed reference). Second, this convexity is known to give regularity to the solution (see \cite{LoMi1}, \cite{LoeperQuiros}). Third, it gives the user some control about choosing a meaningful reference model from which one tries not to depart too much.

We choose as a starting reference model a classic  parametric model, calibrated as well as possible to market prices. As we explain below, to help convergence and smoothness, we propose to update iteratively the reference model: In each iteration, the reference model is taken as the output of the previous iteration. The reference model is also used as the starting point of the gradient descent. This ensures that the reference model is chosen if it matches the calibration constraints.

To illustrate the above methodology, we consider the CEV (constant elasticity of variance) model of \cite{cox1975notes,cox1996constant} under which the underlying, $(S_t)_{t\geq 0}$, follows 
\begin{equation}\label{eq: CEV}
\diff S_t = r_t S_t\diff t + \sigma(t,S_t)S_t\diff W^1_t,
\end{equation}
with $\sigma(t,S_t) = \sigma S_t^{\gamma - 1}$, where $\sigma,\gamma \geq 0$ are constants. 
In the sequel, we write $Z_t=\log(S_t)$ and work with the log-price process. The short rate $(r_t)_{t\geq 0}$, is modelled using the Vasicek model of \cite{vasicek1977equilibrium}, and follows 
\begin{equation}\label{eq: hull-white}
\diff r_t = a(b-r_t)\diff t +\sigma_r\diff W^2_t,
\end{equation}
with
\begin{equation}\label{eq: CEV-HW cov}
\diff\langle W^1,W^2\rangle_t = \rho \diff t.
\end{equation}

The constants $a,b,\sigma_r$ are chosen strictly positive, with $a$ the speed of mean reversion, $b$ the mean, and $\sigma_r$ the volatility of the short rate. 

All the modelling is done under the risk neutral (pricing) measure and the initial market values $S_0, r_0$ are given. We note that our methods extend easily to non-trivial initial distribution of $S_0, r_0$ which could be useful for forward-starting models, or a multi-period calibration.  
 We fix the parameters of our reference model as $(\overline{\sigma},\overline{\sigma}_r,\overline{\rho},\overline{a},\overline{b})$ and denote 
\begin{equation} 
 X_t = (X^1_t,X^2_t)=(Z_t,r_t).
 \end{equation} The drift coefficient $\overline\alpha(t,X_t)$ and the diffusion coefficient $\overline \beta(t,X_t) =\frac{d\langle X\rangle_t}{dt}$ of the process $X$ under the reference model are thus given by:
\begin{equation}\label{eq: CEV-HW char}
(\overline{\alpha}(t,X_t),\overline{\beta}(t,X_t))=\left(\begin{bmatrix} 	r_t-\frac{1}{2}(\exp(Z_t))^{2(\overline{\gamma}-1)}\\ \overline{a}\left(\overline{b} - r_t\right)\end{bmatrix},\begin{bmatrix}\overline{\sigma}^2\exp(Z_t)^{2(\overline{\gamma}-1)}&\bar{\rho}\bar{\sigma}\bar{\sigma}_r\exp(Z_t)^{\overline{\gamma}-1}\\\bar{\rho}\bar{\sigma}\bar{\sigma}_r\exp(Z_t)^{\overline{\gamma}-1} &\bar{\sigma}_r^2\end{bmatrix}\right).
\end{equation}
Note that on the other hand there is no restriction  on the parameters of the calibrating model, apart from risk neutrality and finiteness of the cost function. Our cost function being convex,  see \eqref{eq: cost function} below, combined with the classical Markovian mimicking arguments, see \cite{gyongy1986mimicking,brunick2013mimicking} and \cite[Lemma~3.2]{JLO23calibration}, imply that with no loss of generality we can restrict our attention to Markovian models where $\alpha,\beta$ are functions of $(t,X_t)$. Note that the initial value $X_0\in\IR^2$ is known and hence a potential candidate model is specified simply via a choice of $(\alpha,\beta): [0,T]\times \IR^2 \to \IR^2\times\IS^2_+$, where $\IS^2_+$  are positive definite symmetric $2\times 2$ matrices. Risk neutrality implies $\alpha_1=r-\frac{1}{2}\beta_{11}$. Formally, these observations mean that our candidate model is of the following form, where we take probability measures $\IP$ from the set of measures such that $\IE^{\IP}\left[\int_0^T\!|\alpha(t,X_t)|+|\beta(t,X_t)|\,\diff t\right]<+\infty$, which we denote as $\IP\in\cP^1$, and we let $W^{\IP}$ be a Brownian motion under $\IP$:
\begin{equation}\label{eq: MP SDE}
\begin{cases}
\diff X_t=\alpha(t,X_t)\diff t+(\beta(t,X_t))^{\frac{1}{2}}\diff W_t^{\IP},&0\leq t\leq T,\\
X_0=x_0.&
\end{cases}
\end{equation}
Further, we impose bounds on the diffusion coefficients:
$$
\delta^l_{11}\leq\beta_{11}(t,x)\leq \delta^u_{11}\qquad\delta^l_{22}\leq\beta_{22}(t,x)\leq\delta^u_{22},\qquad \text{for all }(t,x)\in [0,T]\times\IR^2.
$$
This is desirable from the point of numerical stability, as well as being justifiable from the financial modelling point of view. 
The drift and diffusion then take values in the following admissible (convex) set, which depends on the interest state variable:
\begin{equation}\label{eq: gamma joint}
\Gamma_{\mathrm{joint}}(r)\coloneqq\left\{(\alpha,\beta)\in\IR^2\times\IS^2_+\: :\:\alpha_1=r-\frac{1}{2}\beta_{11},\:\delta_{11}^l\leq\beta_{11}\leq\delta^u_{11},\:\delta^l_{22}\leq\beta_{22}\leq\delta^u_{22}\right\}.
\end{equation}
Define $||\cdot||_{\mathrm{Fro}}$ as the Frobenius norm, which is given by $||M||_{\mathrm{Fro}}=\sqrt{\sum_{i,j}M_{i,j}^2}$.
We consider the following cost function $F$ 
\begin{equation}\label{eq: cost function}
F(\alpha,\beta)=\begin{cases} ||\alpha-\overline{\alpha}||^2_2+||\beta-\overline{\beta}||_{\mathrm{Fro}}^2,&\mathrm{if}\:(\alpha,\beta)\in\Gamma_{\mathrm{joint}},\\
+\infty,&\mathrm{otherwise}.\end{cases}
\end{equation}
For two square matrices $A,B$, denote the matrix inner product by $A:B=\mathrm{tr}(A^{\intercal}B)$. The Legendre-Fenchel transform of $F$ (see \cite{rockafellar1970convex}) with respect to $(\alpha,\beta)$ is defined by
\begin{equation}\label{eq: LF of cost function}
F^*(a,b)\coloneqq\sup_{\substack{\alpha\in\IR^2,\beta\in\IS_+^2\\\ \delta^l_{11}\leq\beta_{11}\leq\delta^u_{11}\\ \delta^l_{22}\leq\beta_{22}\leq\delta^u_{22}}}\{\alpha\cdot a + \beta : b-F(\alpha,\beta)\}.
\end{equation}
We consider market  data given as prices of $n$ European options on $X$. 
Their maturities and payoffs are denoted respectively by $\tau=(\tau_1,\dots,\tau_n)\in (0,T)^n$ and $G(x)=(G_1(x),\dots,G_n(x))$, and for technical reasons we approximate the call payoffs so that $G_i :\IR^2\to\IR$ is continuous and bounded for all $i=1,\dots,n$. The market price of these options are given by $u=(u_1,\dots,u_n)$ so given a candidate model $\IP\in\cP^1$, our market constraints are:
$$
\IE^{\IP}\left[e^{-\int_{0}^{\tau_i}\! r_s\,\diff s}G_i(X_{\tau_i})\right]=u_i,\quad i=1,\dots, n.
$$

We denote models $\IP\in\cP^1$ such that the above constraints are satisfied as $\IP\in\cP^1_{\mathrm{loc}}$. Note that in practice, we will only consider call options on $S$ and caps on interest rates. 
Given these constraints, we obtain the reference model (\ref{eq: CEV-HW char}) via a least squares parametric calibration. Specifically, let $\tilde{u}(\overline{\sigma},\overline{\sigma}_r,\overline{\gamma},\overline{\rho},\overline{a},\overline{b})\in\IR^n$ denote the prices of the  $n$ options attained our model \eqref{eq: CEV}--\eqref{eq: CEV-HW cov} with parameters $(\overline{\sigma},\overline{\sigma}_r,\overline{\gamma},\overline{\rho},\overline{a},\overline{b})$. The reference model is obtained by numerically solving:
\begin{equation}
\min_{\overline{\sigma},\overline{\sigma}_r,\overline{\gamma},\overline{\rho},\overline{a},\overline{b}}\frac{1}{n}\sum_{i=1}^n\left(\tilde{u}_i(\overline{\sigma},\overline{\sigma}_r,\overline{\gamma},\overline{\rho},\overline{a},\overline{b}) - u_i\right)^2.
\end{equation}
Where the minimisation is taken over $\overline{\sigma},\overline{\sigma}_r,\overline{a},\overline{b}>0$ and $\overline{\rho}\in [-1,1]$. To minimise this, we simply compute the price of the options given the parameters $(\overline{\sigma},\overline{\sigma}_r,\overline{\gamma},\overline{\rho},\overline{a},\overline{b})$ using the Feynman-Kac formula, then numerically compute the gradients with respect to the parameters using a central difference approximation, and then apply the L-BFGS algorithm of \cite{liu1989limited}. We ran the optimisation algorithm to a first order error of $1\times 10^{-3}$, which corresponded to an error of implied volatilities less than $10^{-2}$ in the SPX call options, and of $10^{-1}$ in the short rate cap options.
\begin{table}[H]
\centering
\begin{tabular}{|l|l|l|p{0.4\linewidth}|}
\hline
Parameter & Initial Value & Parametrically Calibrated Value & Interpretation\\
\hline
$\overline{\sigma}$ & 0.4 & 0.4115 & Volatility scaling of the CEV model\\
$\overline{\gamma}$ & 0.9 & 0.9362 & Power law in the CEV model\\
$\overline{\rho}$ & -0.2 & -0.2037 & Instantaneous correlation between short rate and log-stock\\
$\overline{\sigma}_r$ & 0.03 & 0.0232 & Volatility of the Vasicek model\\
$\overline{a}$ & 0.5 & 0.0156 & Speed of mean reversion in the Vasicek model\\
$\overline{b}$ & 0.03 & 0.2852 & Mean to which Vasicek model reverts\\
\hline
\end{tabular}
\caption{Reference Model Parameters}
\label{table: ref params}
\end{table}
Now that the reference model is set, we can  formulate the SOT-Calibration problem of \cite{JLO23calibration}. We seek to find 
$$ \inf_{\IP\in\cP^1_{\mathrm{loc}}}\IE^{\IP}\left[\int_0^T e^{-\int_0^t r_s \diff s} F(\alpha(t,X_t),\beta(t,X_t))\diff t\right].$$
 This can rephrased in terms of $(\alpha,\beta)$ and the discounted density $\rho$, i.e., if for $\IP\in\cP^1_{\mathrm{loc}}$ that $\tilde\rho(t,x)=\IP(X_t\in dx)$ is the density of $X_t$ then $\rho(t,x)=\IE^{\IP}[e^{-\int_0^t r_s\,\diff s}\tilde\rho(t,X_t)|\sigma(X_t)]$. We formalise this as follows:
\begin{lemma}[Lemma~3.3 of \cite{JLO23calibration}]
\label{lem:discountedFP}
Let $\IP\in \cP^1_{loc}$ so that $X$ solves \eqref{eq: MP SDE}. Let $\eta_{t,x}(\cdot)$ be the law of $\int_0^t r_s\,\diff s$ conditional on $X_t=x$. Define the `discounted density'
\begin{equation}\label{eq: dd substitution}
\rho(t,x)\coloneqq\left(\int_{\IR}e^{-y}\eta_{t,x}(\mathrm{d}y)\right)\bar\rho(t,x)=D(t,x)\bar\rho(t,x),\qquad (t,x)\in[0,T]\times\IR^2. 
\end{equation}
Then ${\rho}$ solves the `discounted' version of the Fokker-Planck equation for $(t,x)\in[0,T]\times\IR^2$:
\begin{equation}\label{eq: discount fp}
\partial_t{\rho}(t,x)+\nabla_x\cdot\left(\alpha(t,x){\rho}(t,x)\right)-\frac{1}{2}\nabla_x^2:\left(\beta(t,x){\rho}(t,x)\right)+r{\rho}(t,x)=0. 
\end{equation}
\end{lemma}
\begin{problem}[Primal Problem]\label{prob: primal}
\begin{equation}\label{eq: objective function}
V=\inf_{\rho,\alpha,\beta}\int_0^T\int_{\IR^2}\!F(\alpha(t,x),\beta(t,x))\rho(t,\diff x)\diff t,
\end{equation}
where the infimum is taken over $(\rho,\alpha,\beta)\in C([0,T];\mathcal{M}(\IR^2))\times L^1(\diff\rho_t\diff t;\IR^2)\times L^1(\diff\rho_t\diff t;\IS^2)$ subject to:	
\begin{align*}
\partial_t\rho(t,x)+\nabla_x(\rho(t,x)\alpha(t,x))-\frac{1}{2}\nabla^2_x:(\rho(t,x)\beta(t,x))+r\rho(t,x)&=0,\quad (t,x)\in[0,T]\times\IR^2,\\
\int_{\IR^d}\!G_i(x)\rho(\tau_i,\diff x)&=u_i,\quad\text{for }i=1,\dots,n,\\
\rho(0,\cdot)&=\delta_{X_0}.
\end{align*}
\end{problem}
The above Fokker-Planck equation is understood in the sense of distributions.  The duality for the above problem was established in \cite[Theorem 3.5]{JLO23calibration} and the dual problem is given by:
\begin{thm}[Dual Problem]\label{prob: dual}
$$
V=\sup_{\lambda,\phi}\lambda\cdot u - \phi^{\lambda}(0,Z_0,r_0),
$$
where the supremum is taken over $\lambda\in\IR^n$ and $\phi^{\lambda}$ is the (unique, discontinuous in time) viscosity solution to the HJB equation:
\begin{align}
\partial_t\phi^{\lambda}+\sup_{\substack{\alpha_2\in\IR,\beta\in\IS^2_+,\\ \delta^l_{11}\leq\beta_{11}\leq\delta^u_{11},\\ \delta^l_{22}\leq\beta_{22}\leq\delta^u_{22}}}\biggl\{&\left(r-\frac{1}{2}\beta_{11}\right)\partial_{z}\phi^{\lambda}+\alpha_2\partial_{r}\phi^{\lambda}+\frac{1}{2}\beta_{11}\partial^2_{zz}\phi^{\lambda}+\frac{1}{2}\beta_{22}\partial^2_{rr}\phi^{\lambda}+\beta_{12}\partial^2_{zr}\phi^{\lambda}\notag\\
&-||\alpha-\overline{\alpha}||^2_2-||\beta-\overline{\beta}||^2_{\mathrm{Fro}}\biggr\}-r\phi^{\lambda}+\sum_{i=1}^n\lambda_iG_i(x)\delta_{\tau_i}=0,\qquad \text{for }(t,x)\in[0,T]\times\IR^2,\label{eq: HJB JC}
\end{align}
with terminal condition $\phi^{\lambda}(T,\cdot)=0$. If $V$ is finite, then the infimum in Problem~\ref{prob: primal} is attained. If the sup is attained for some $\lambda^*$ and corresponding $\phi^*$ solving (\ref{eq: HJB JC}), then the optimisers $(\alpha^*,\beta^*)$ are given by
\begin{equation}\label{eq: form of optimisers}
(\alpha_t^*,\beta_t^*)=\nabla F^*\left(\nabla_x\phi^*(t,\cdot),\frac{1}{2}\nabla^2_x\phi^*(t,\cdot)\right),\qquad \diff\rho_t\diff t\text{ - almost everywhere.}
\end{equation}
\end{thm}
Note that we use the definition of viscosity solution given in \cite{guo2022calibration,guo2022joint,JLO23calibration}, which adapts the classical notion of viscosity solutions from \cite{lions1983optimal} to include the jump discontinuities and also includes the terminal condition $\phi(T,\cdot) = 0$.

\section{Solving the Semimartingale Optimal Transport Calibration Problem}
\subsection{Numerical Solutions for the Dual Problem}\label{sec: numerics}
Our numerical approach focuses on solving the dual problem. Specifically, we solve the HJB equation (\ref{eq: HJB JC}) by following \cite{JLO23calibration}, which is an adaptation of the methods presented in \cite{guo2022calibration} and \cite{guo2022joint}. We use a discretisation on a uniform $100\times 100$ spatial grid of $[7.6,8.8]\times [0,5]$ for the log-stock and rescaled interest rates, and partition the time interval into days, so that $\diff t = \frac{1}{365}$. We discretise the HJB equation using an implicit finite difference method, with central difference approximations for the spatial derivatives. We choose a boundary far away enough such that the boundary conditions have less of an effect on the HJB equation solution, and our boundary conditions are such that the second derivative of $\phi$ does not change with time between each calibrating option. That is, for all $x$ on the boundary of our computational domain, and for a subsequence of the calibrating option maturity times $(\tau_{i_k})_{k=1,\dots,m}$ such that for $k=1,\dots,m$ all $\tau_{i_k}$ are distinct, and with $\tau_{i_0}=0$,
$$
\nabla^2_x\phi(t,x) = \nabla^2_x\phi(\tau_i,x),\quad\text{for }t\in (\tau_{i_{k-1}},\tau_{i_k}],\:k=1,\dots,m.
$$
(\ref{eq: HJB JC}) is a fully nonlinear parabolic PDE and we use a policy iteration method (see \cite{ma2017unconditionally}) as in \cite{JLO23calibration} to solve it. In our policy iteration procedure to approximate $\alpha^*$ and $\beta^*$, we compute the finite difference approximations of $\nabla\phi$ and $\nabla^2\phi$ using $\phi$ from the previous iteration. The $\phi$ used in the initial iteration is $\phi$ from the previous time step (that is $\phi_{\mathrm{init}}(t_n,\cdot)\coloneqq\phi(t_{n+1},\cdot)$). With this approximated $(\alpha^*,\beta^*)$, we can apply an implicit finite differences method to one step of the HJB equation to obtain the new value of $\phi$. This iteration is repeated until some specified tolerance $\epsilon_2$ is reached such that $||\phi_{\mathrm{old}}(t_n,\cdot)-\phi_{\mathrm{new}}(t_n,\cdot)||_{\infty}<\epsilon_2$. In comparison to \cite{JLO23calibration}, we are now approximating all of the coefficients, so the iteration takes more steps and thus is less efficient at computing $\alpha^*$ and $\beta^*$. The jump discontinuities are handled by simply adding $\lambda_iG_i(x)$ to the $\phi_{\mathrm{init}}$ when we reach the timestep corresponding to $\tau_i$. Once we have solved the HJB equation at all time steps, we then solve the linearised model pricing PDE (with coefficients $\alpha^*$ and $\beta^*$) via the ADI method to generate the model prices.  

Having solved the HJB equation for a fixed $\lambda$, we turn our attention to solving Problem~\ref{prob: dual} and finding the optimal $\lambda^*$. First, we observe that we can speed up the optimisation routine by providing a formula for the gradients. This formula is obtained in the same way as \cite{guo2022joint}, but with the discounting appearing as a result of the $-r\phi$ term in the HJB equation and the Feynman-Kac formula.

\begin{lemma}\label{lem: gradients}
Suppose Problem~\ref{prob: primal} is admissible, and define the dual objective function as
\begin{equation}\label{eq: dual objective}
L(\lambda) = \lambda\cdot u - \phi(0,X_0).
\end{equation}
Then the gradients of the dual objective function are given by
\begin{equation}\label{eq: dual gradients}
\partial_{\lambda_i}L(\lambda) = u_i - \IE^{\IP}\left[e^{-\int_0^{\tau_i}\!r_s\,\diff s}G_i(X_{\tau_i})\right].
\end{equation}
\end{lemma}
Our initialisation is $\lambda=0$ since if the reference model is already calibrated, then that will immediately return $||\nabla_{\lambda}L(\lambda)||_{\infty}<\epsilon_1$. Given a guess $\lambda$, we solve the HJB equation (\ref{eq: HJB JC}). From the solution of \eqref{eq: HJB JC}, we compute the diffusion coefficients $(\alpha^*(t,x),\beta^*(t,x))$ from \eqref{eq: form of optimisers}. For each instrument we compute the model price by solving
\begin{equation}\label{eq: pricing PDE}
\begin{cases} \partial_t\psi(t,x)+\alpha^*(t,x)\cdot\nabla_x\psi(t,x)+\frac{1}{2}\beta^*(t,x):\nabla^2_x\psi(t,x)-r\psi(t,x)=0,&\quad (t,x)\in [0,\mathcal{T})\times\IR^2,\\
\psi(\mathcal{T},x)=\mathcal{G}(x),&\quad x\in\IR^2.\end{cases}
\end{equation}

We can then compute the gradient $\nabla_{\lambda}L(\lambda)$ corresponding to the difference between the model prices and  and market prices, and finally use the L-BFGS algorithm to update $\lambda$.

\begin{algorithm}[H]\label{algo: dual}
\caption{Policy iteration algorithm.}
\DontPrintSemicolon
\SetNoFillComment
\KwData{Input an initial $\lambda$ and market prices $u_i$.}
\KwResult{Calibrated model prices, optimal drift and diffusion}
\While{$||\nabla_{\lambda}L(\lambda)||_{\infty}>\epsilon_1$}{
\tcc{Solve the HJB equation backwards in time}
\For{$k=N-1,\dots,0$}{
\tcc{Terminal Conditions - adding $\lambda$ multiplied by the payoff}
\If{$t_{k+1}=\tau_i$ for some $i=1,\dots,n$}{
$\phi_{t_{k+1}}\gets\phi_{t_{k+1}}+\sum_{i=1}^n\lambda_iG_i\indic{\{t_{k+1}=\tau_i\}}$\
}
\tcc{Policy iteration to approximate the optimal characteristics}
$\phi^{\mathrm{new}}_{t_k}\gets\phi_{t_{k+1}}$\tcp*[r]{Approximate using previous time step}
\While{$||\phi^{\mathrm{new}}_{t_k}-\phi^{\mathrm{old}}_{t_k}||>\epsilon_2$}{
$\phi^{\mathrm{old}}_{t_k}\gets\phi^{\mathrm{new}}_{t_k}$\tcp*[r]{Store the old value of $\phi$}\
Approximate $(\alpha^*,\beta^*)$ from $\phi^{\mathrm{old}}_{t_k}$. \tcp*[r]{Use old values to approximate optimal characteristics}\
Plug $(\alpha^*,\beta^*)$ into (\ref{eq: HJB JC}) to remove the supremum and solve using one step of an implicit finite difference method, and set the solution to $\phi^{\mathrm{new}}_{t_k}$.
}
$\phi_{t_k}\gets\phi^{\mathrm{new}}_{t_k}$\tcp*[r]{Save the solution once the $\phi$ has converged to the optimal solution}
}
\tcc{Computing the model prices and gradients}
Compute the model prices by solving the pricing PDE (\ref{eq: pricing PDE}) using the ADI method.\;
Compute the gradients (\ref{eq: dual gradients}).\;
Use the L-BFGS algorithm to update $\lambda$.
}
\end{algorithm} 

\subsection{Approximating the optimisers in \eqref{eq: form of optimisers}}
In the step 9 of Algorithm \ref{algo: dual} we have to Approximate $(\alpha^*,\beta^*)$ from $\phi^{\mathrm{old}}_{t_k}$. This is a key step and it needs to be done efficiently. It is possible to derive an analytic formula for $(\alpha^*,\beta^*)$ but this proves to have many subcases and to involve solving quartic equations. In consequence, this method is computationally costly. In our numerical experiments it proved far slower than the alternative which we now describe. 

\cite{corrias1996fast} provides a numerical scheme to evaluate the Legendre-Fenchel transform in $O(N^d\log N)$ time for $N$ discretisation points and $d$ dimensions. We remark as well that \cite{lucet1997faster} provides a faster version in $O(N^d)$ time, however it is difficult to separate the $\beta$ terms while maintaining positive semi-definiteness. Since (\ref{eq: form of optimisers}) requires the evaluation of the spatial derivatives at all points of the grid, we cannot directly apply the methods to fully take advantage of the faster computational speed. However, we use the observation in equation (2.1) of \cite{corrias1996fast}, which allows us to decompose the maximisation component-wise, and apply the idea of evaluating the transform on bounded intervals for computational ease. 
\begin{lemma}[Approximating the Optimal Coefficients]\label{lem: optimisers joint}
Define $(\alpha^*,\beta^*)$ by:
\begin{align*}
\alpha^*_1(t,z,r)&= r - \frac{1}{2}\beta^*_{11}(t,z,r),\\
\alpha^*_2(t,z,r)&= \overline{\alpha}_2(t,z,r) + \frac{1}{2}\partial_r\phi(t,z,r),\\
\beta^*_{11}(t,z,r)&= \begin{cases}\overline{\beta}_{11}(t,z,r) + \frac{1}{5}(\partial^2_{zz}\phi(t,z,r)-\partial_z\phi(t,z,r)),&\text{ when }\overline{\beta}_{11} + \frac{1}{5}(\partial^2_{zz}\phi-\partial_z\phi)\in[\delta_{11}^{l},\delta_{11}^u],\\
\delta_{11}^l,&\text{ when }\overline{\beta}_{11} + \frac{1}{5}(\partial^2_{zz}\phi-\partial_z\phi)<\delta_{11}^l,\\
\delta_{11}^u,&\text{ when }\overline{\beta}_{11} + \frac{1}{5}(\partial^2_{zz}\phi-\partial_z\phi)>\delta_{11}^u,\end{cases}\\
\beta_{22}^*(t,z,r) &= \begin{cases}\overline{\beta}_{22}(t,z,r)+\frac{1}{4}\partial^2_{rr}\phi(t,z,r),&\text{ when }\overline{\beta}_{22} + \frac{1}{4}\partial^2_{rr}\phi\in[\delta_{22}^{l},\delta_{22}^u],\\
\delta_{22}^l,&\text{ when }\overline{\beta}_{22} + \frac{1}{4}\partial^2_{rr}\phi<\delta_{22}^l,\\
\delta_{22}^u,&\text{ when }\overline{\beta}_{22} + \frac{1}{4}\partial^2_{rr}\phi>\delta_{22}^u,\end{cases}\\
\beta_{12}^*(t,z,r)&=\begin{cases}\overline{\beta}_{12}(t,z,r)+ \frac{1}{4}\partial^2_{zr}\phi(t,z,r),&\text{ when } \overline{\beta}_{12} + \frac{1}{4}\partial^2_{zr}\phi\in\left[-\sqrt{\beta^*_{11}\beta^*_{22}},\sqrt{\beta^*_{11}\beta^*_{22}}\right],\\
-\sqrt{\beta^*_{11}\beta^*_{22}},&\text{ when }\overline{\beta}_{12} + \frac{1}{4}\partial^2_{zr}\phi< -\sqrt{\beta^*_{11}\beta^*_{22}},\\
\sqrt{\beta^*_{11}\beta^*_{22}},&\text{ when }\overline{\beta}_{12} + \frac{1}{4}\partial^2_{zr}\phi>\sqrt{\beta^*_{11}\beta^*_{22}},\end{cases}
\end{align*}
and $\beta_{21}^*=\beta_{12}^*$. 
The matrix $\beta^*$ is positive semidefinite and whenever 
\begin{equation}\label{eq: beta12 bounds}
-\sqrt{\beta_{11}^*\beta_{22}^*}<\beta_{12}^*<\sqrt{\beta_{11}^*\beta_{22}^*}
\end{equation}
holds, then $(\alpha^*,\beta^*)$ are the optimisers in \eqref{eq: form of optimisers}. 
\end{lemma}
The proof is given in Appendix \ref{sec:appendix opt}. It proceeds by solving (\ref{eq: form of optimisers}) sequentially: we first solve for $\beta_{11}$ and $\beta_{22}$ subject to their bounds. This is easy to do as the expression is quadratic in each variable. We then solve for $\beta_{12}$ and use it to enforce the condition of positive semi-definiteness on the matrix $\beta$. If \eqref{eq: beta12 bounds} holds then the positive semi-definitiveness condition is not binding and the procedure returns the optimizer. Otherwise, we set $\beta_{12}^*=\pm \sqrt{\beta_{11}^*\beta_{22}^*}$ to ensure $\beta^*$ is positive semi-definite but this means our approximation may differ from a global search over all positive semi-definite matrices $\beta$. However, in all of our numerical experiments, \eqref{eq: beta12 bounds} in fact holds everywhere so that, in practice, Lemma \ref{lem: optimisers joint} provides the optimisers in \eqref{eq: form of optimisers} but at a fraction of the computational cost involved in solving \eqref{eq: form of optimisers} analytically as a constrained optimization \eqref{eq: LF cost fct}.

\subsection{Renormalisation and Reference Model Iteration}
As in \cite{JLO23calibration}, in order to bring the short rate to the same order as the log-stock for finite difference approximation stability reasons, we perform the constant rescaling $r_t\mapsto R r_t$ where we choose $R=100$. In addition, we rescale the calibrating option prices and their payoffs by their vegas computed from their Black-Scholes implied volatility. This not only helps the stability of the numerical method, but also converts pricing errors into implied volatility errors since the vega represents how much the option price will change as the volatility changes by $1\%$.

If Algorithm~\ref{algo: dual} is directly applied, then often it will output a model with spiky drift and diffusion surfaces which is undesirable. This is a result of our cost function penalising deviations from a reference model, which results in larger local changes to the surfaces, but with the rest of the overall surface being closer to the reference model. Moreover, instabilities from the $\partial_{xx}\phi$ and $\partial_{rr}\phi$ terms in $\beta_{11}$ and $\beta_{22}$ at maturity at the strikes of our options when adding in the jump discontinuities are unavoidable. In addition, sometimes the L-BFGS algorithm will get stuck and unable to progress. To avoid all of this and improve the convergence to our calibrated model, we apply a ``reference model iteration'' technique as in \cite{guo2022joint}. That is, once our optimisation routine finishes, we apply an interpolation and smoothing technique to the output model drift and diffusion terms, and then set those terms to our new reference model and re-run the calibration. The final output model is not smoothed since this would no longer be a calibrated model. This not only speeds up convergence, but also decreased sharp peaks that can arise in our calibrated local volatility surfaces otherwise.

\section{Numerical Results: Market Data Example}

We test our calibration procedure on market data\footnote{Data obtained from a Bloomberg terminal at the Sa\"{i}d Business school in Oxford on 23/05/2022.}. We used the one month LIBOR as a proxy for the short rate, and obtained the implied volatility and prices of the following options:
\begin{itemize}
\item 10 calls on the SPX with expiry 19/08/2022, 
\item 6 caps on the one month LIBOR with notional \$10,000,000 and expiry 23/08/2022, 
\item 10 calls on the SPX with expiry 18/11/2022,
\item 6 caps on the one month LIBOR with notional \$10,000,000 and expiry 23/11/2022.
\end{itemize}
We have a total of $n=32$ options with payoff functions $G_i(X_{\tau_i}) = (\exp(X^1_{\tau_i}) - K_i)^+$ for the calls and $G_i(X_{\tau_i}) = N\tau_i(X^2_{\tau_i} - K_i)^+$ for the caps, where $N$ is the notional value.
\begin{figure}[H]
\centering
	\subfigure[]{\includegraphics[width=0.45\textwidth]{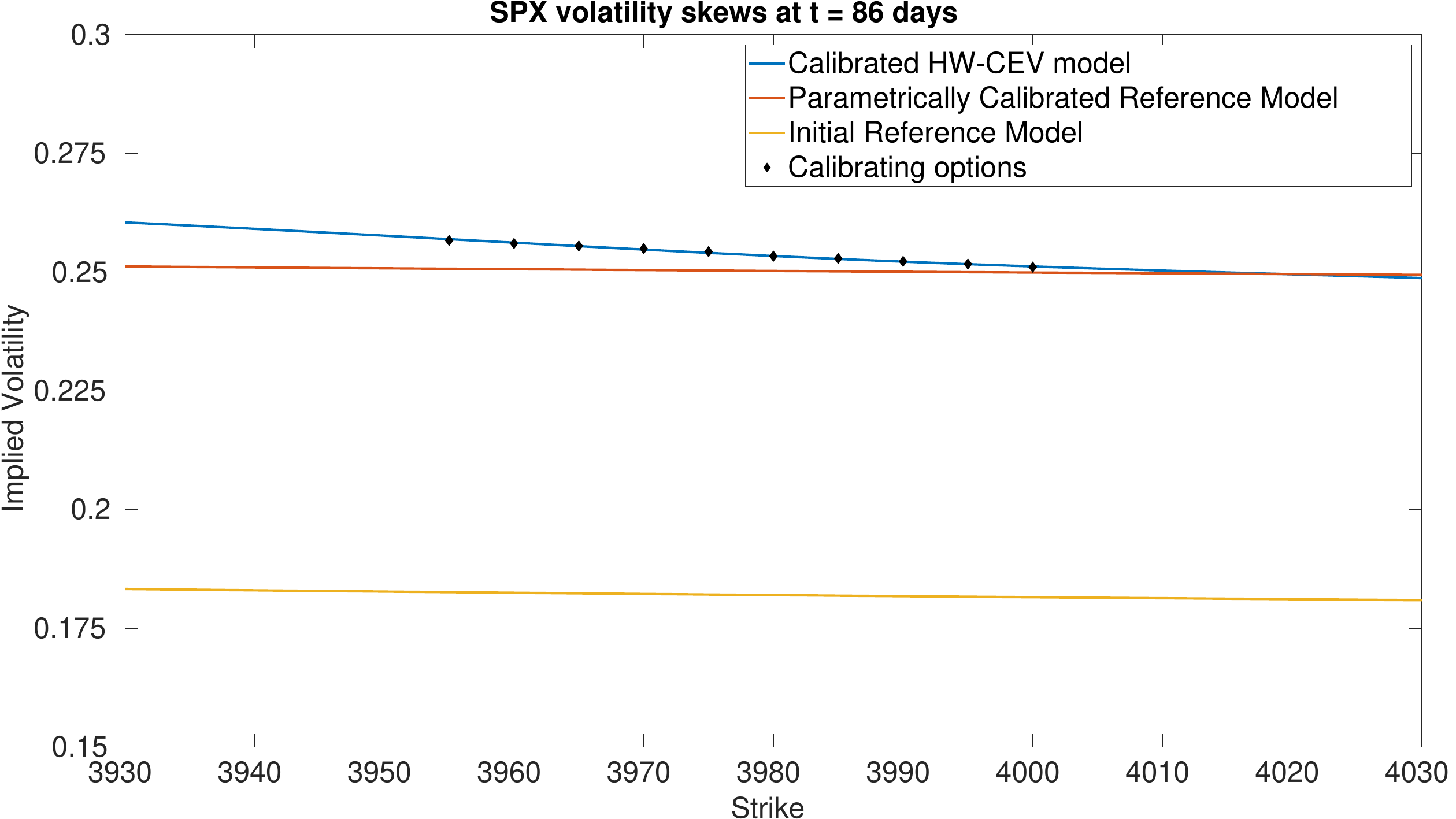} }
	\subfigure[]{\includegraphics[width=0.45\textwidth]{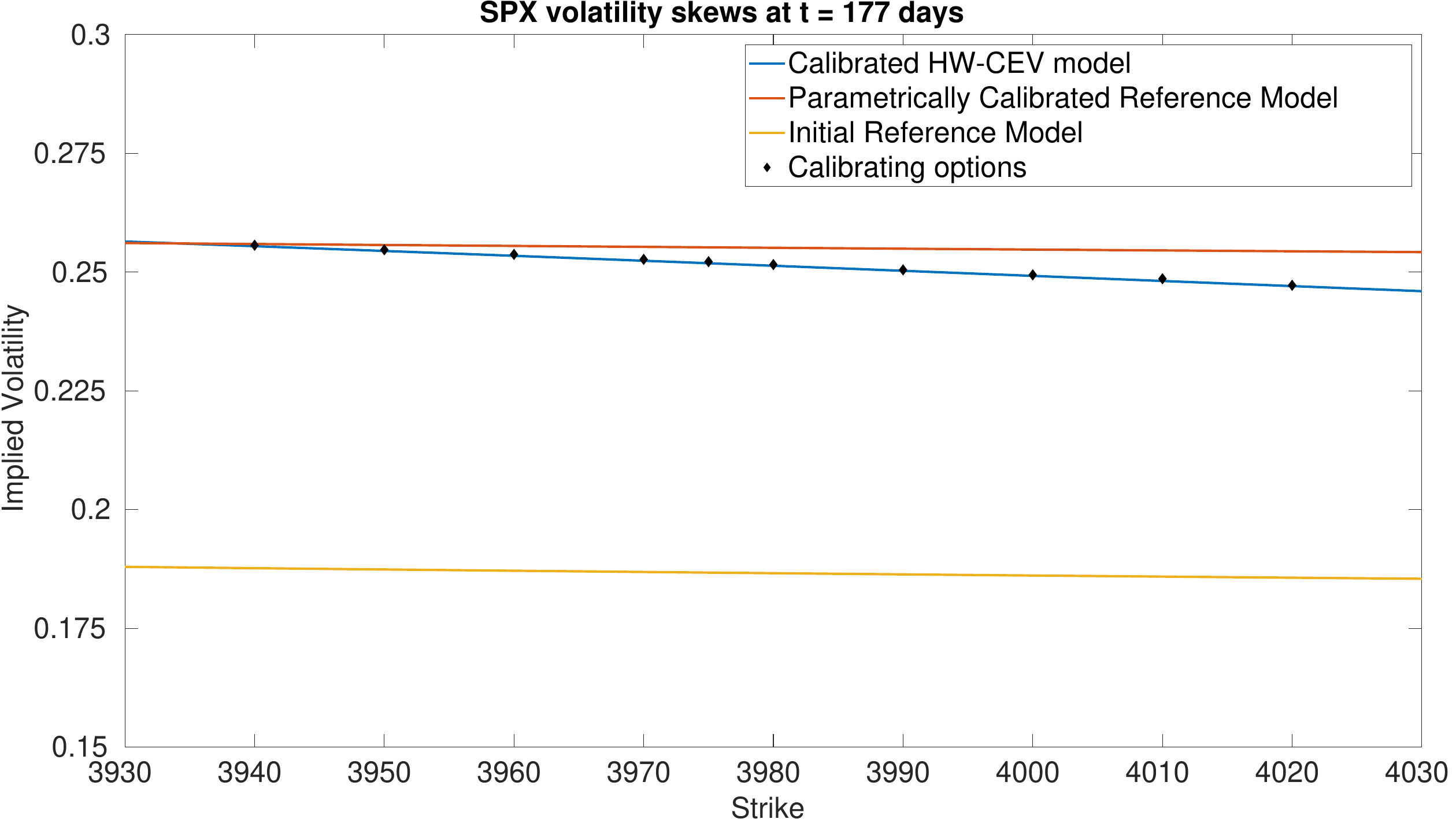} }
	\caption{Implied volatility skews of the SPX calibrating options under the reference model and calibrated HW-CEV model.}
\end{figure}
\begin{figure}[H]
\centering
	\subfigure[]{\includegraphics[width=0.45\textwidth]{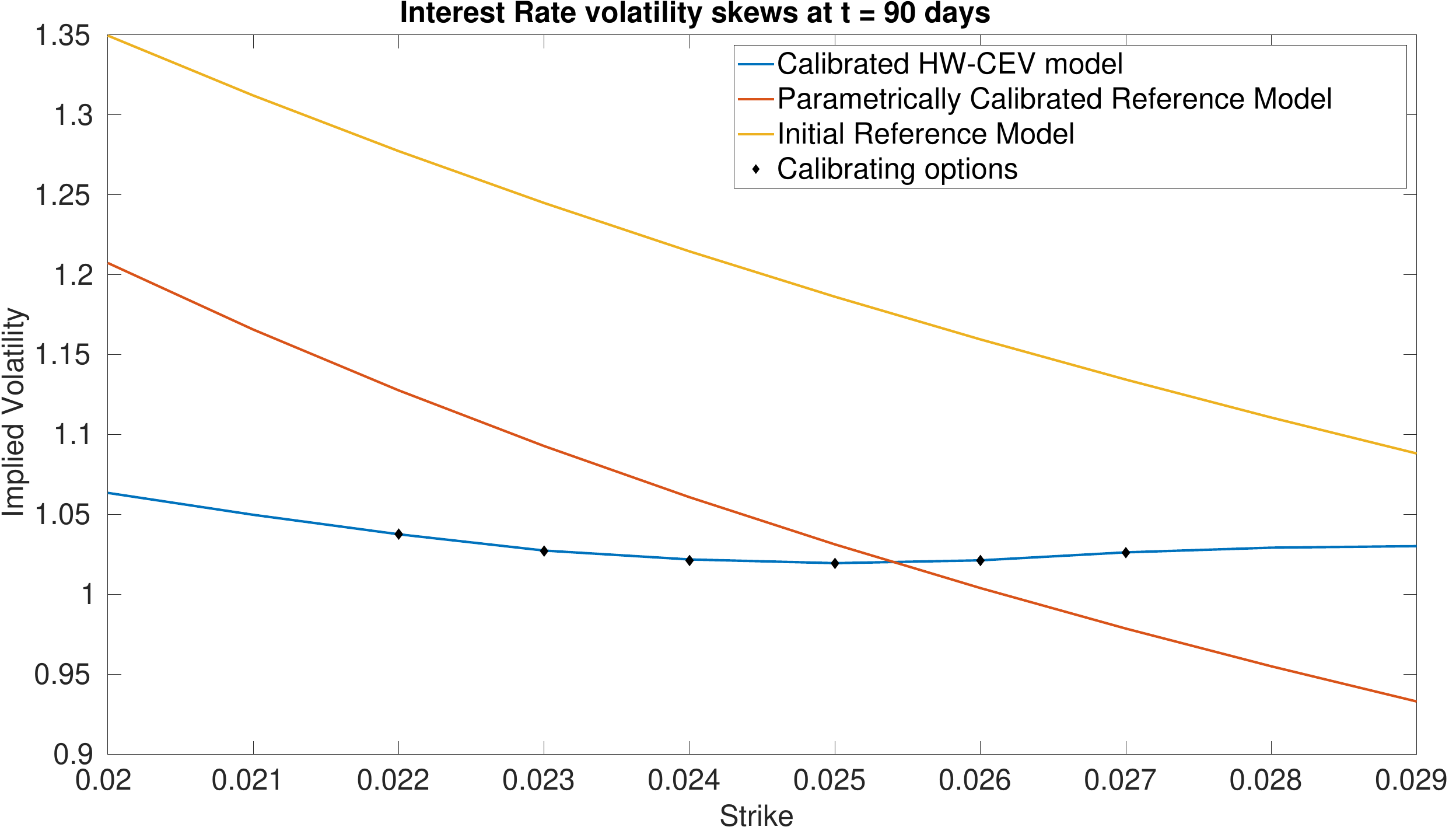} }
	\subfigure[]{\includegraphics[width=0.45\textwidth]{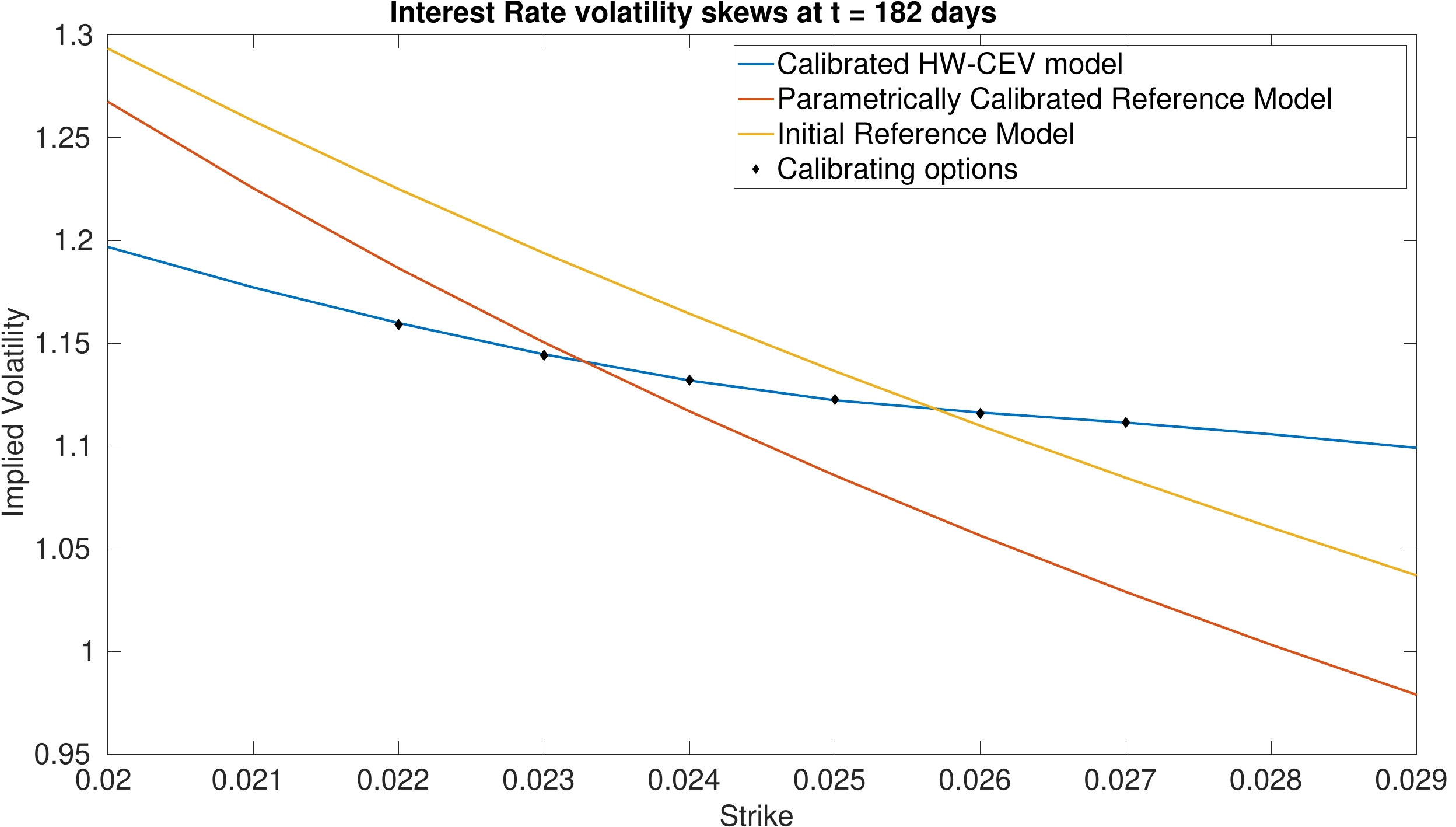} }
	\caption{Implied volatility skews of the Short Rate calibrating options under the reference model and calibrated HW-CEV model.}
\end{figure}
\begin{figure}[H]
\centering
	\subfigure[]{\includegraphics[width=0.35\textwidth]{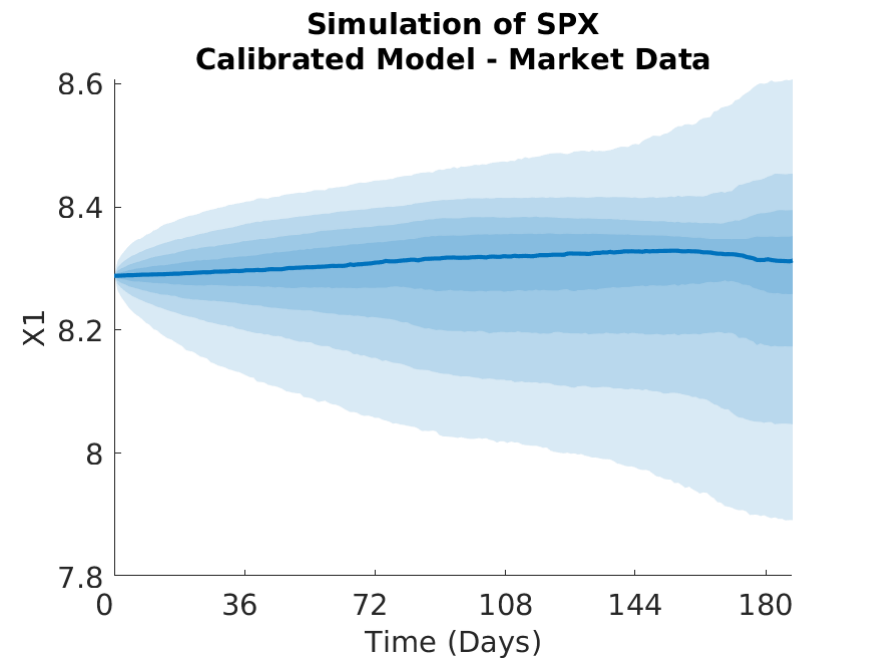}}
	\subfigure[]{\includegraphics[width=0.35\textwidth]{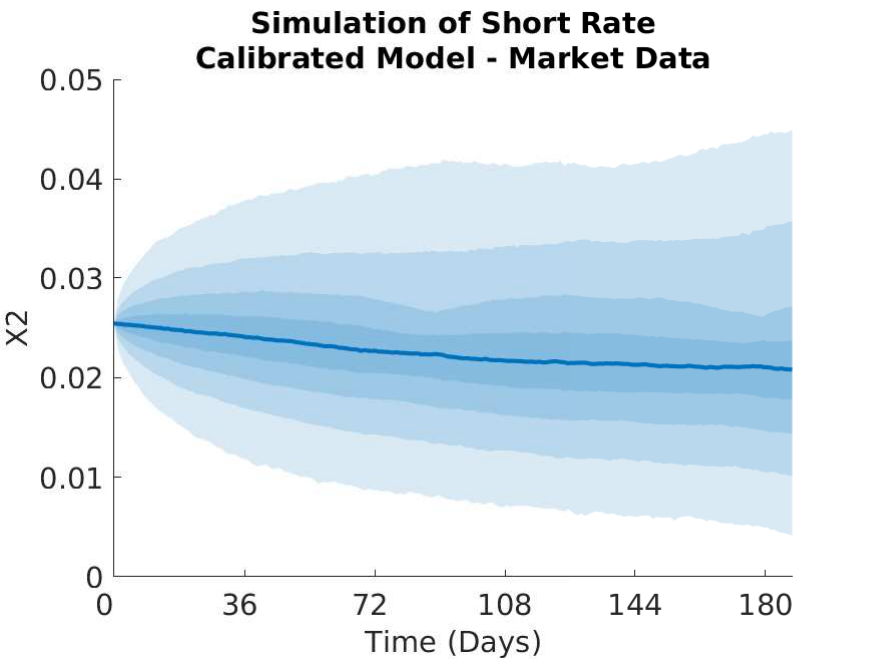}}
	\caption{Monte Carlo simulations of (a) the log-SPX and (b) the short rate in the SOT calibrated model.}
\label{fig:MC SPX}
\end{figure}
We display the optimisation parameters, and the bounds for $\beta_{11}$ and $\beta_{22}$ below. It is difficult to choose the bounds since too relaxed a bound will result in large values for $\beta_{11}$ and $\beta_{22}$ and also result in numerical instabilities for $\beta_{12}$; whereas too tight a bound will severely slow down the calibration. Even with these bounds chosen, we still saw some instabilities in $\beta_{11}$ on the maturity dates of the Calls, and in $\beta_{22}$ on the maturity dates of the Caps. Imposing the bounds reduced these instabilities significantly, but did not remove them entirely.
\begin{table}[H]
\centering
\begin{tabular}{|l|l|p{0.6\linewidth}|}
\hline
Parameter & Value & Interpretation\\
\hline
$\epsilon_1$ & $1\times 10^{-3}$ & Tolerance for the difference in model and observed IV\\
$\epsilon_2$ & $1\times 10^{-8}$ & Tolerance in the policy iteration for $\phi$\\
$\delta_{11}^l$ & $0.01$ & Lower bound for $\beta_{11}$\\
$\delta_{11}^u$ & $0.5$ & Upper bound for $\beta_{11}$\\
$\delta_{22}^l$ & $2\times 10^{-4}$ & Lower bound for $\beta_{22}$\\
$\delta_{22}^u$ & $1\times 10^{-3}$ & Upper bound for $\beta_{22}$\\
\hline
\end{tabular}
\caption{Optimisation Parameters and Bounds}
\end{table}
\subsubsection{Plots of Drift and Diffusion Surfaces}
\begin{figure}[H]
\centering
\includegraphics[width=0.8\textwidth]{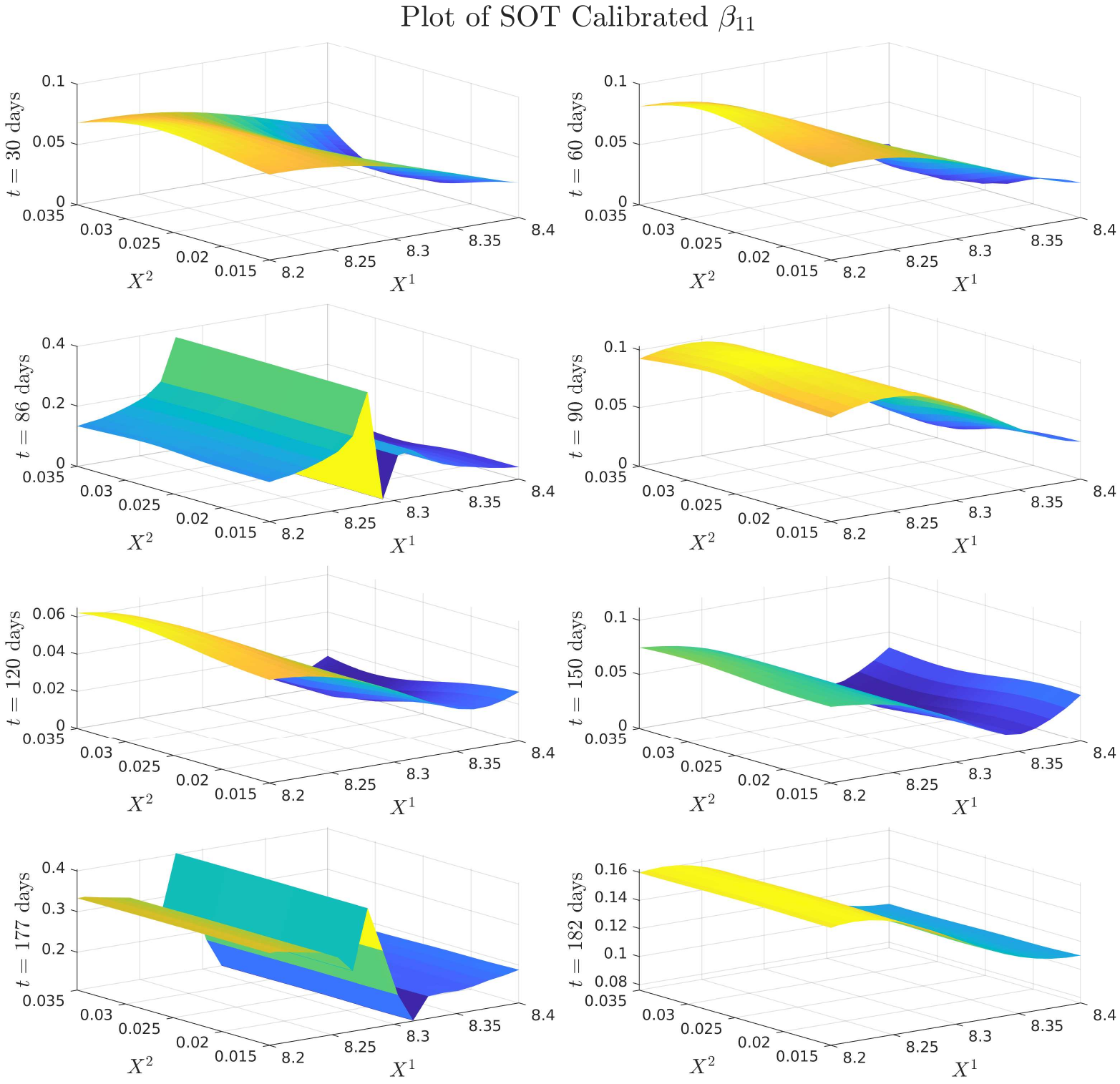} 
\caption{Plots of SOT-calibrated $\beta_{11}$, the volatility of the log-stock.}
\end{figure}
\begin{figure}[H]
\centering
\includegraphics[scale=0.4]{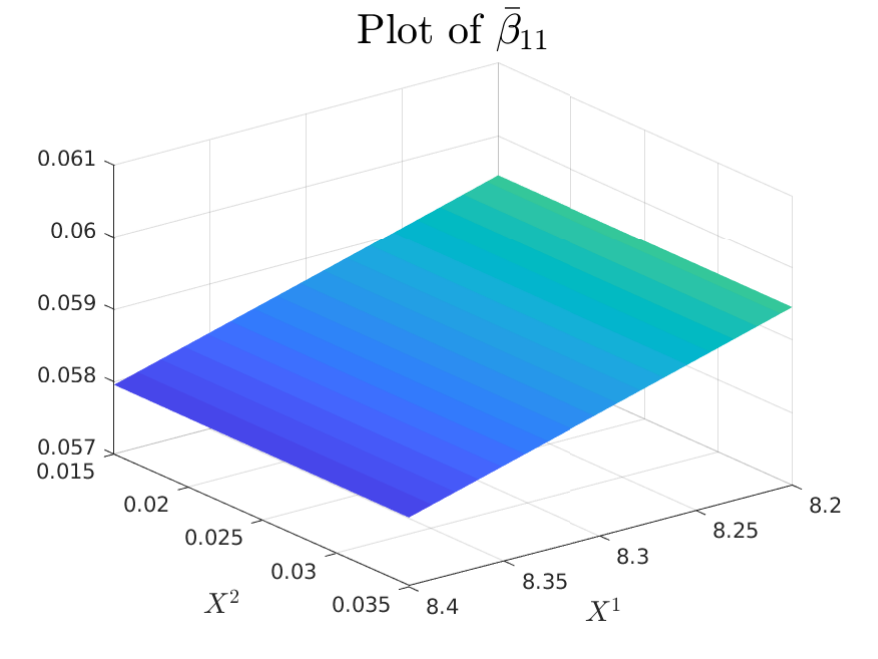}
\caption{Plot of parametrically calibrated reference $\overline{\beta}_{11}$. Note that this is taken to be time homogeneous.}
\end{figure}
\begin{figure}[H]
\centering
\includegraphics[width=0.8\textwidth]{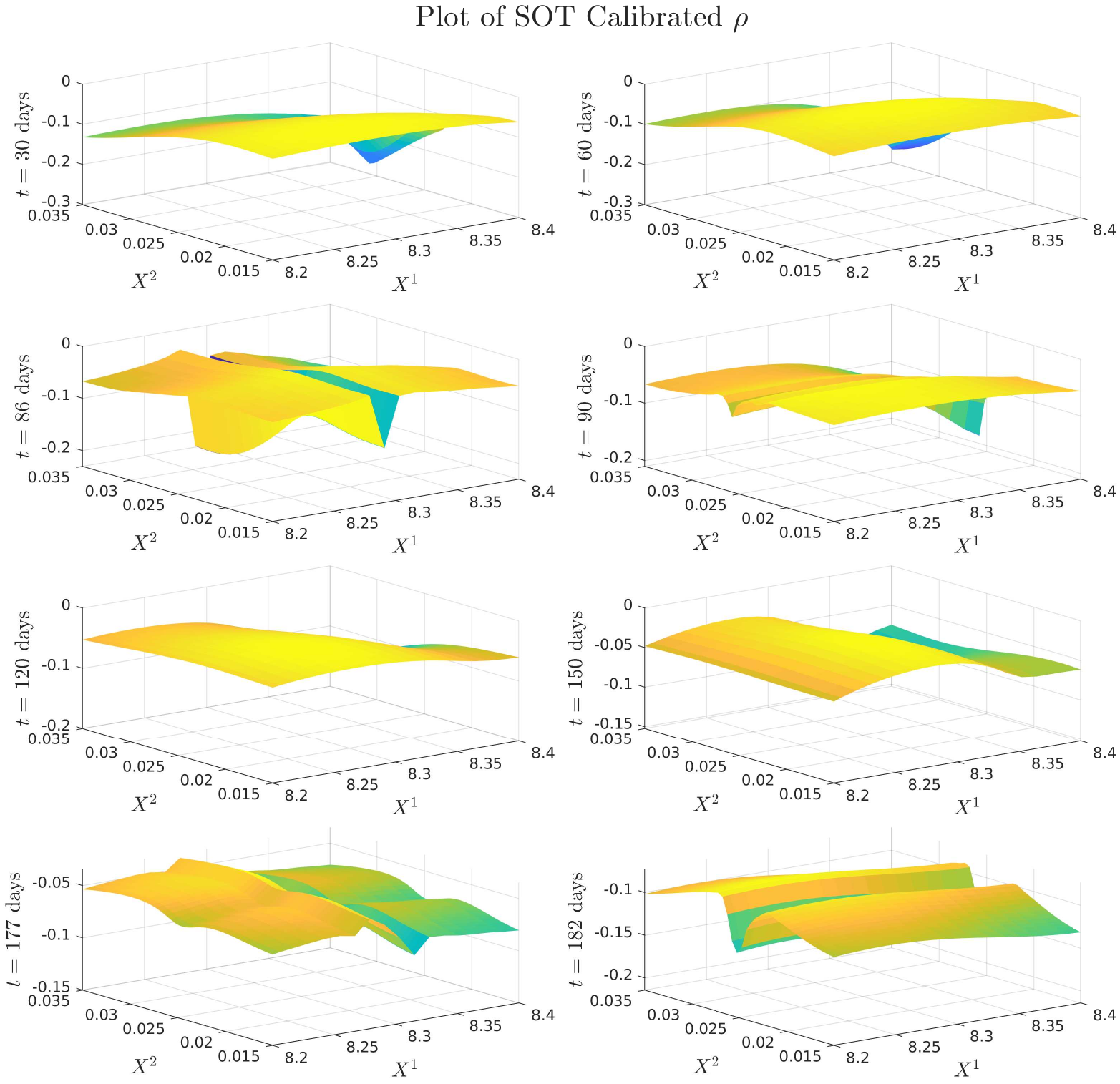} 
\caption{Plots of SOT calibrated correlation $\rho = \frac{\beta_{12}}{\sqrt{\beta_{11}\beta_{22}}}$. The $\beta_{12}$ and $\beta_{22}$ used here have the scaling undone in order to obtain the actual correlation between $\log(S_t)$ and $r_t$. The parametrically calibrated reference model takes constant value $\overline{\rho} = -0.2037$.}
\end{figure}
\begin{figure}[H]
\centering
\includegraphics[width=0.8\textwidth]{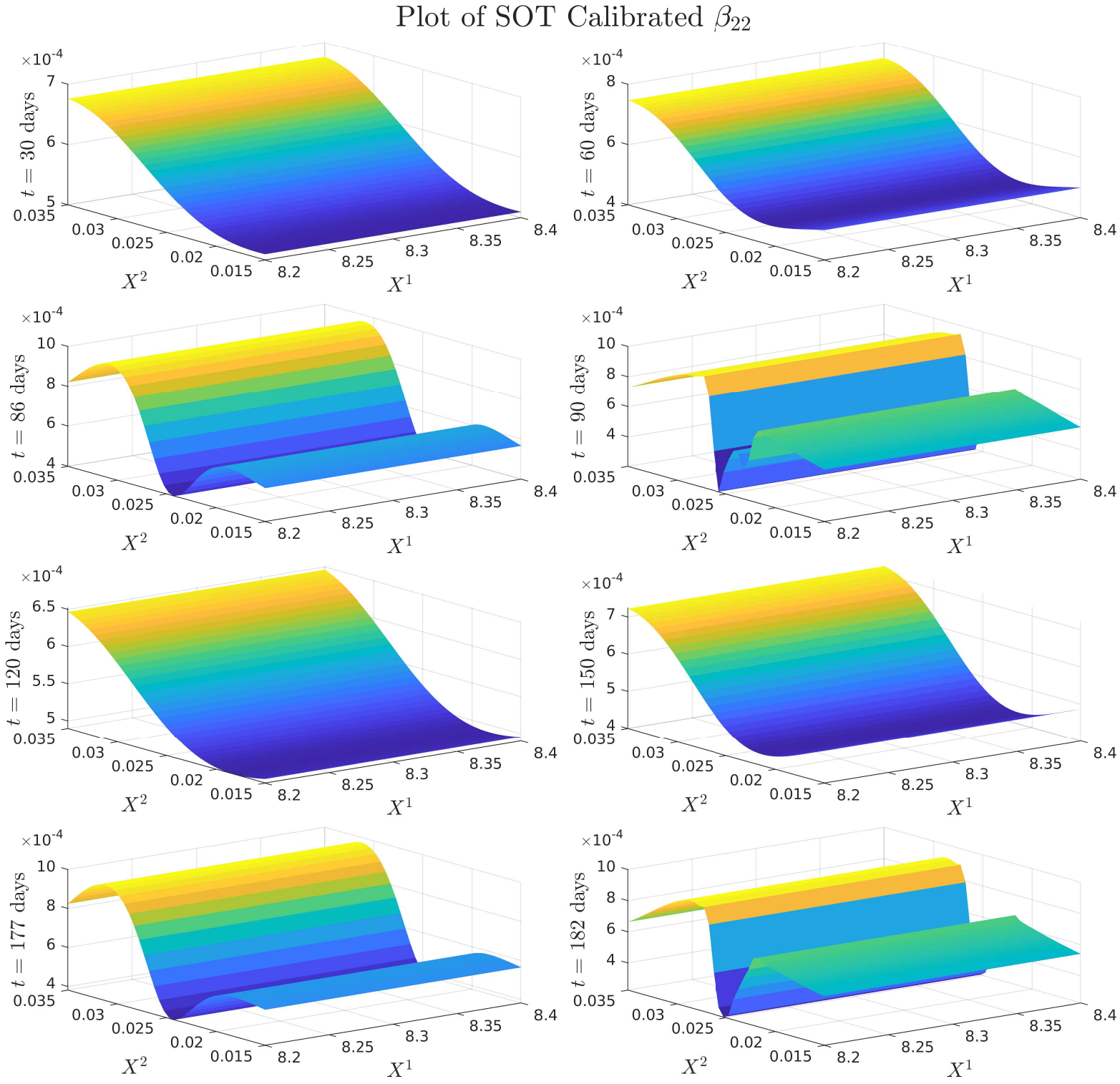} 
\caption{Plots of SOT calibrated $\beta_{22}$, the volatility of the short rate. Note that the $\beta_{22}$ used have the scaling undone to represent the actual volatility of the short rate. The parametrically calibrated reference model takes constant value $\overline{\beta}_{22}=5.3753\times 10^{-4}$.}
\end{figure}
\begin{figure}[H]
\centering
\includegraphics[width=0.8\textwidth]{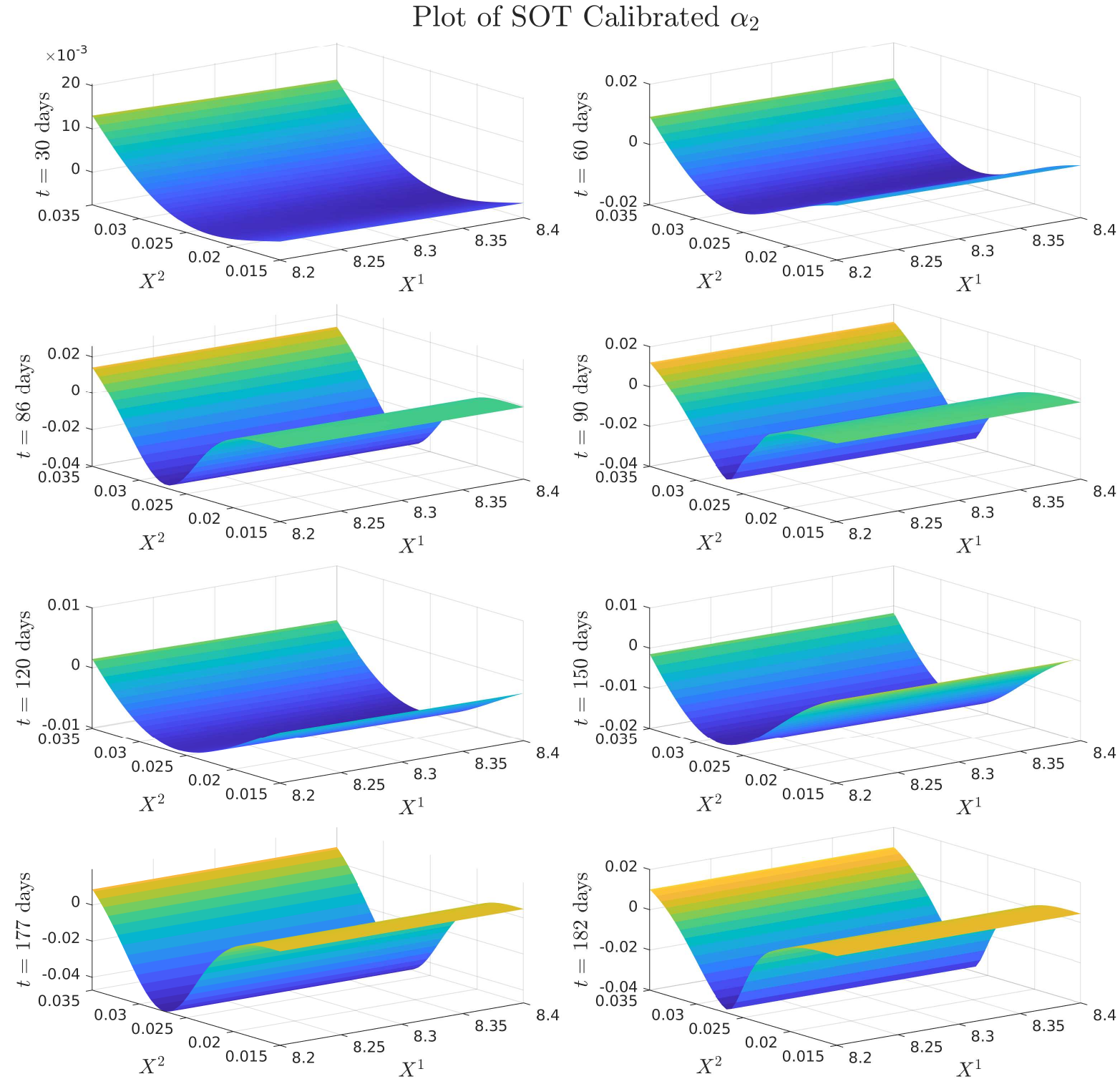} 
\caption{Plots of SOT calibrated $\alpha_{2}$. Note that the $\alpha_2$ used have the scaling undone to represent the actual drift of the short rate.}
\end{figure}
\begin{figure}[H]
\centering
\includegraphics[scale=0.4]{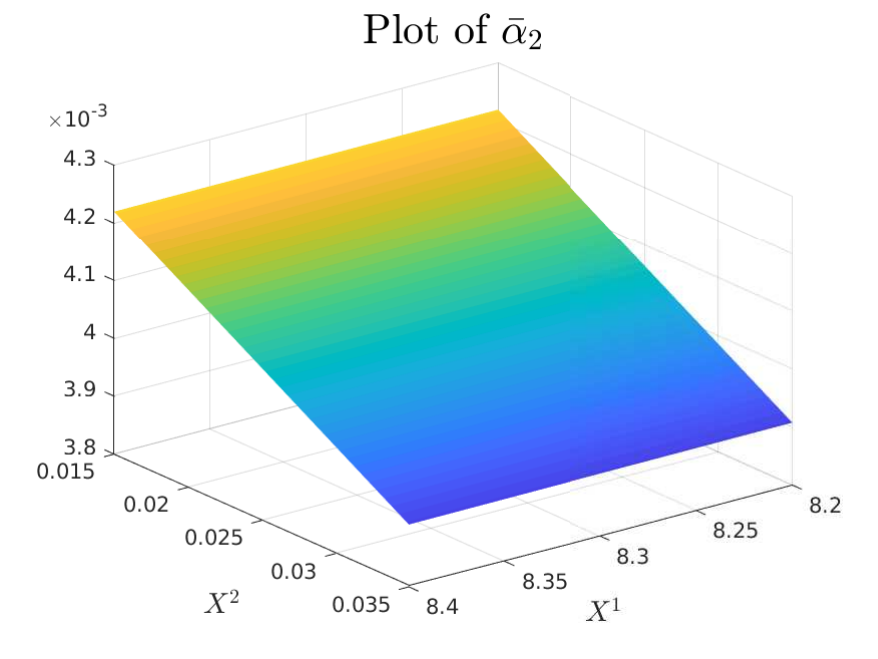}
\caption{Plot of parametrically calibrated reference $\overline{\alpha}_{2}$. Note that this is taken to be time homogeneous.}
\end{figure}
\section{Numerical results: a comparison of three SOT Calibration approaches}
Above we develop a joint calibration method for European options on a stock and on interest rates. In \cite{JLO23calibration} a sequential approach was proposed in which the interest rates model was calibrated first and frozen for the subsequent calibration of a local volatility model for the stock prices. This was done under rigid restrictions on the structure of the correlation -- it was conjectured this would speed up the numerics. We want to now compare the two approaches numerically, as well as introduce and compare a third approach which also proceeds sequentially but relaxes the assumptions on the correlation.  

We test the methods on simulated data. We use the generating model as the reference model for the short rate. In the two cases where the interest rate is frozen, the surface plots of $\alpha_2$ and $\beta_{22}$ will remain the same as the reference model, however $\alpha_2$ and $\beta_{22}$ may be perturbed in the joint calibration case so we show all surface plots.
Specifically, we take a Hull-White model for the interest rate, see \cite{hull1990pricing, hull1994branching}, and local volatility dynamics for the log-price:
\begin{align}
\diff Z_t &= r_t - \frac{1}{2}\sigma^2(t,Z_t,r_t)+\sigma(t,Z_t,r_t)\diff W^1_t\nonumber\\
\diff r_t &= (b(t)-a r_t)\diff t +\sigma_r\diff W^2_t\label{eq:model SDE}\\
\diff\langle W^1,W^2\rangle_t&=\rho(t,Z_t,r_t)\diff t\nonumber
\end{align}
Where $a,\sigma_r(t)>0$ are constants and $b(\cdot)$ is a function of time, calibrated so that the dynamics of $(r_t)_{0\leq t\leq T}$ match the market data (e.g., suitable interest rates caps and floors). Both $a$ and $\sigma_r$ being positive constants is not a particularly restrictive constraint as remarked in \cite{brigo2007interest,hull1995note}. Note that $b(t)$ will therefore need to be calibrated to fit the term structure of interest rates seen in the market. Our reference model for the local volatility of the log-stock will be a CEV model, as above.

The difference in the calibration approach for all three methods is what process(es) are being calibrated, or equivalently the cost function employed, which then also leads to different formulae for the optimal coefficients. We review briefly the sequential method used in \cite{JLO23calibration} and define its extension, the ``full sequential'' method. We then provide numerical results comparing the three approaches. 
\subsection{Sequential Calibration}
We fix a reference function $\rho_{\mathrm{ref}}(t,Z_t,r_t)\in\IR^2$ and restrict correlation to the following representation:
\begin{equation}\label{eq: convexity adjustment}
\rho(t,Z_t,r_t)=\frac{\sigma_r(t)}{\sigma(t,Z_t,r_t)}\rho_{\mathrm{ref}}(t,Z_t,r_t),\quad\text{for}\:t\in[0,T].
\end{equation}
We then set:
\begin{equation}\label{eq: gamma sequential case}
\Gamma_{\mathrm{seq}}(t,r)=\left\{(\alpha,\beta)\in\IR^2\times\IS^2_+\: :\:\alpha_1=r-\frac{1}{2}\beta_{11},\:\alpha_2=(b(t)-ar),\:\beta_{12}=\beta_{21}=\rho_{\mathrm{ref}}\sigma_r^2,\:\beta_{22}=\sigma_r^2\right\},
\end{equation} 
We also require the inequality $\rho_{\mathrm{ref}}(t,Z_t,r_t)^2\sigma_r(t)^2\leq\sigma^2(t,Z_t,r_t)$ to keep $\rho(t,Z_t,r_t)\in [-1,1]$ as a correlation also. Since $r_t$ is on a much lower scale to $Z_t$, we have that $\sigma_r^2\ll\sigma^2$, so this condition is not financially restrictive. To enforce this condition, we will define a convex function $H:\IR\times\IR^+\times\IR\to\IR\cup\{ +\infty\}$ with a parameter $p>2$:
$$
H(x,\bar{x},s)\coloneqq\begin{cases}
(p-1)\left(\frac{x-s}{\bar{x}-s}\right)^{1+p}+(p+1)\left(\frac{x-s}{\bar{x}-s}\right)^{1-p}-2p,&\text{if}\: x,\bar{x}>s,\\
+\infty,&\text{otherwise}.
\end{cases}
$$
The coefficients of each term ensure that $H(x,\bar{x},s)$ is minimised over $x$ at $x=\bar{x}$ with $\min_x H(x,\bar{x},s)=0$. We fix a reference local volatility function $\bar{\sigma}^2=\bar{\sigma}^2(t,Z_t,r_t)$ that represents the desired model. Since we wish to penalise deviations away from our reference volatility $\overline{\sigma}^2$, we define our cost function $F$ as follows:
\begin{equation}\label{eq: sequential cost function}
F_{\mathrm{seq}}(t,r,\alpha,\beta)=\begin{cases}
H(\beta_{11},\bar{\sigma}^2,\rho_{\mathrm{ref}}^2\sigma_r^2),&\text{if}\:(\alpha,\beta)\in\Gamma_{\mathrm{seq}}(t,r),\\
+\infty,&\text{otherwise}.
\end{cases}
\end{equation}
With this cost function and the reference model fixed by $\Gamma_{\mathrm{seq}}$ in (\ref{eq: gamma sequential case}), our dual formulation becomes
\begin{problem}[Sequential Dual Formulation]\label{prob: seq dual}
$$
V=\sup_{\lambda}\lambda\cdot u-\phi^{\lambda}(0,Z_0,r_0).
$$
Where $\phi^{\lambda} = \phi(t,z,r)$ solves the HJB equation:
\begin{align}
\sum_{i=1}^n\lambda_iG_i(x)\delta_{\tau_i}+\partial_t\phi+&\sup_{\beta_{11}}\biggl(\left(r-\frac{1}{2}\beta_{11}\right)\partial_{z}\phi+(b(t)-ar)\partial_{r}\phi+\frac{1}{2}\beta_{11}\partial^2_{zz}
\phi+\rho_{\mathrm{ref}}\sigma_r^2\partial^2_{zr}\phi+\frac{1}{2}\sigma^2_r\partial^2_{rr}\phi-r\phi\notag\\&\qquad\qquad-H(\beta_{11},\bar{\sigma}^2,\rho_{\mathrm{ref}}^2\sigma_r^2)\biggr)=0,\quad (t,z,r)\in [0,T]\times\IR^2.\label{eq: HJB seq cal}
\end{align}
\end{problem}
\begin{lemma}[Analytic Formula for the Optimal Characteristic in Sequential Calibration]\label{lem: optimal sequential}
The optimal characteristic in the HJB equation (\ref{eq: HJB seq cal}), $\beta_{11}^*$ is given by
\begin{equation}\label{eq: optimal beta11}
\beta_{11}^*(t,z,r) = \rho_{\mathrm{ref}}^2\sigma_r^2+\left(\frac{(\bar{\sigma}^2-\rho_{\mathrm{ref}}^2\sigma_r^2)^p(\phi_{zz}-\phi_z)}{4(p^2-1)}+\frac{1}{2}\sqrt{\left(\frac{(\bar{\sigma}^2-\rho_{\mathrm{ref}}^2\sigma_r^2)^p(\phi_{zz}-\phi_z)}{2(p^2-1)}\right)^2+4(\bar{\sigma}^2-\rho_{\mathrm{ref}}^2\sigma_r^2)^{2p}}\right)^{\frac{1}{p}}.
\end{equation}
\end{lemma}
We use $p=4$ in our numerical experiments, see Table \ref{tab:paramters_compare}.
\subsection{``Full Sequential'' Calibration}\label{sect: fs calib}
We introduce the notion of ``full sequential'' calibration, in which the correlation is treated as a free parameter rather than prescribed via \eqref{eq: convexity adjustment}. Our set $\Gamma$ now becomes:
\begin{equation}\label{eq: gamma full sequential case}
\Gamma_{\text{full seq}}(t,r)=\left\{(\alpha,\beta)\in\IR^2\times\IS^2_+\: :\:\alpha_1=r-\frac{1}{2}\beta_{11},\:\alpha_2=(b(t)-ar),\:\beta_{22}=\sigma_r^2\right\}.
\end{equation} 
We use the same cost function as in joint calibration, however since we have from the definition of $\Gamma_{\text{full seq}}$ that $\alpha_2=\overline{\alpha}_2$ and $\beta_{22}=\overline{\beta}_{22}$, and moreover that $\alpha_1-\overline{\alpha}_1=\frac{1}{2}(\overline{\beta}_{11}-\beta_{11})$, our cost function will simplify to:
\begin{equation}\label{eq: full sequential cost function}
F_{\text{full seq}}(t,r,\alpha,\beta)=\begin{cases}\frac{5}{4}(\beta_{11}-\overline{\beta}_{11})^2+2(\beta_{12}-\overline{\beta}_{12})^2,&\text{if }(\alpha,\beta)\in\Gamma_{\text{full seq}}(t,r),\\
+\infty,&\text{otherwise.}\end{cases}
\end{equation}
\begin{problem}[Full Sequential Dual Formulation]\label{prob: full seq dual}
$$
V=\sup_{\lambda}\lambda\cdot u-\phi^{\lambda}(0,Z_0,r_0).
$$
Where $\phi^{\lambda} = \phi(t,z,r)$ solves the HJB equation:
\begin{align}
\sum_{i=1}^n\lambda_iG_i(x)\delta_{\tau_i}+\partial_t\phi+&\sup_{\substack{\beta_{11},\beta_{12},\\ \beta\in\IS^2_+}}\biggl(\left(r-\frac{1}{2}\beta_{11}\right)\partial_{z}\phi+(b(t)-ar)\partial_{r}\phi+\frac{1}{2}\beta_{11}\partial^2_{zz}\phi+\beta_{12}\partial^2_{zr}\phi+\frac{1}{2}\sigma^2_r\partial^2_{rr}\phi-r\phi\notag\\&\qquad\qquad-F_{\text{full seq}}(t,r,\alpha,\beta)\biggr)=0,\quad (t,z,r)\in [0,T]\times\IR^2.\label{eq: HJB Full seq cal}
\end{align}
\end{problem}
In a similar way to Lemma~\ref{lem: optimisers joint}, we obtain an approximation of the optimal $\beta_{11}$ and $\beta_{12}$ by first optimising over $\beta_{11}$, and then over $\beta_{12}$ applying the positive semidefinite constraint in this variable given $\beta_{11}^*$. 
\begin{lemma}\label{lem: optimal full sequential}
Let $\beta_{22}^*=\sigma_r^2$ and define $\beta_{11}^*$ and $\beta_{12}^*=\beta_{21}^*$ by:
\begin{align*}
\beta_{11}^*(t,z,r)&=\begin{cases}\overline{\beta}_{11}(t,z,r) + \frac{1}{5}(\partial^2_{zz}\phi(t,z,r)-\partial_z\phi(t,z,r)),&\text{ when } \overline{\beta}_{11} + \frac{1}{5}(\partial^2_{zz}\phi-\partial_z\phi)\in[\delta_{11}^l,\delta_{11}^u]\\
\delta_{11}^l,&\text{ when }\overline{\beta}_{11} + \frac{1}{5}(\partial^2_{zz}\phi-\partial_z\phi)<\delta_{11}^l,\\
\delta_{11}^u,&\text{ when }\overline{\beta}_{11} + \frac{1}{5}(\partial^2_{zz}\phi-\partial_z\phi)>\delta_{11}^u,\end{cases}\\
\beta_{12}^*(t,z,r)&=\begin{cases}\overline{\beta}_{12}(t,z,r)+ \frac{1}{4}\partial^2_{zr}\phi(t,z,r),&\text{ when } \overline{\beta}_{12} + \frac{1}{4}\partial^2_{zr}\phi\in\left[-\sigma_r\sqrt{\beta^*_{11}},\sigma_r\sqrt{\beta^*_{11}}\right],\\
-\sigma_r\sqrt{\beta^*_{11}},&\text{ when }\overline{\beta}_{12} + \frac{1}{4}\partial^2_{zr}\phi< -\sigma_r\sqrt{\beta^*_{11}},\\
\sigma_r\sqrt{\beta^*_{11}},&\text{ when }\overline{\beta}_{12} + \frac{1}{4}\partial^2_{zr}\phi>\sigma_r\sqrt{\beta^*_{11}}.\end{cases}
\end{align*}
Then $\beta^*$ is a positive semi-definite matrix and whenever 
$$
-\sigma_r\sqrt{\beta^*_{11}} < \beta^*_{12} < \sigma_r\sqrt{\beta^*_{11}}
$$
then $\beta^*$ is the optimizer in \eqref{eq: HJB Full seq cal}. 
\end{lemma}
\subsection{Numerical Results}
We solve Problem~\ref{prob: seq dual} and Problem~\ref{prob: full seq dual} using the same methods as in Section \ref{sec: numerics}. We provide a table of parameters used in the generating and reference models for our problem. 
\begin{table}[H]
\centering
\begin{tabular}{|l|l|p{0.7\linewidth}|}
\hline
\multicolumn{3}{|l|}{Hull-White CEV Model}\\
\hline
Parameter & Value & Interpretation\\
\hline
$X^1_0$ & $\log(92)$ & Initial log-stock price\\
$X^2_0$ & $0.025\times 100$ & Initial short rate scaled by $R=100$\\
$\epsilon_1$ & $1\times 10^{-4}$ & Tolerance for the difference in scaled model and market implied volatility\\
$\epsilon_2$ & $1\times 10^{-12}$ & Tolerance for the policy iteration approximation of the optimal characteristics\\
$\delta_{11}^l$ & 0.05 & Lower bound of $\beta_{11}$ in full sequential and joint calibration\\
$\delta_{11}^u$ & 1 & Upper bound of $\beta_{11}$ in full sequential and joint calibration\\
$p$ & 4 & Exponent in sequential calibration cost function\\
\hline
$\sigma$ & 0.60 & Volatility scaling of generating CEV model\\
$\gamma$ & 0.95 & Power law in generating CEV model\\
$b(t)$ & $aX^2_0+\frac{\sigma_r^2}{2a}(1-e^{-2at})$ & Initial term structure of Hull-White generating model\\
$a$& 0.05 & Speed of mean reversion of Hull-White generating model\\
$\sigma_r$ & 0.04 & Volatility of Hull-White generating model\\
$\rho$ & $-0.40$  & Instantaneous correlation between short rate and log-stock in generating model\\
\hline
$\overline{\sigma}$ & 0.90 & Volatility scaling of reference CEV model\\
$\overline{\gamma}$ & 0.89 & Power law in reference CEV model\\
$\overline{b}(t)$ & $\overline{a}X^2_0+\frac{\overline{\sigma}_r^2}{2\overline{a}}(1-e^{-2\overline{a}t})$ & Initial term structure of Hull-White reference model\\
$\overline{a}$& 0.05 & Speed of mean reversion of Hull-White reference model\\
$\overline{\sigma}_r$ & 0.04 & Volatility of Hull-White reference model\\
$\overline{\rho}$ & $-0.20$ & Instantaneous correlation between short rate and log-stock in reference model\\
\hline
\end{tabular}
\caption{Parameter values of generating and reference models used in all three methods.}
\label{tab:paramters_compare}
\end{table}

\begin{table}[h]
\centering
\resizebox{\textwidth}{!}{\begin{tabular}{|c|c|c|c|c|c|c|c|c|c|}
\hline
 & &\multicolumn{2}{c|}{Generating Model} & \multicolumn{2}{c|}{Calibrated Model:} & \multicolumn{2}{c|}{Calibrated Model:} & \multicolumn{2}{c|}{Calibrated Model:}\\
& & \multicolumn{2}{c|}{} &\multicolumn{2}{c|}{Sequential} & \multicolumn{2}{c|}{Full Sequential} & \multicolumn{2}{c|}{Joint}\\
\hline
Option Type&Strike&Price&IV&Price&IV&Price&IV&Price&IV\\
\hline
\multirow{6}{*}{\begin{tabular}{c}SPX Call options\\ $t=60$ days\end{tabular}} & 85 & 11.2142 & 0.4825 & 11.2139 & 0.4825 & 11.2139 & 0.4825 & 11.2152 & 0.4826\\
& 92 &7.3755 & 0.4811 & 7.3757 & 0.4811 & 7.3749 & 0.4811 & 7.3750 & 0.4811\\
& 99 &4.6051 & 0.4803 & 4.6045 & 0.4803 & 4.6038 & 0.4802 & 4.6056 & 0.4804\\
& 106 &2.7426 & 0.4799 & 2.7420 & 0.4798 & 2.7419 & 0.4798 & 2.7426 & 0.4799\\
& 113 &1.5667 & 0.4797 & 1.5665 & 0.4797 & 1.5667 & 0.4797 & 1.5673 & 0.4797\\
& 120 &0.8624 & 0.4795 & 0.8623 & 0.4795 & 0.8629 & 0.4796 & 0.8626 & 0.4796\\
\hline
\multirow{6}{*}{\begin{tabular}{c}SPX Call options\\ $t=120$ days\end{tabular}} & 85 &14.0842 & 0.4821 & 14.0839 & 0.4821 & 14.0838 & 0.4821 & 14.0857 & 0.4822\\
& 92 &10.502 & 0.4809 & 10.5021 & 0.4809 & 10.5007 & 0.4808 & 10.5027 & 0.4809\\
& 99 &7.6696 & 0.4797 & 7.6712 & 0.4798 &7.6699 & 0.4798 & 7.6691 & 0.4797\\
& 106 &5.4943 & 0.4785 & 5.4952 & 0.4785 & 5.4936 & 0.4784 & 5.4952 & 0.4785\\
& 113 &3.8607 & 0.4767 & 3.8609 & 0.4768 & 3.8597 & 0.4767 & 3.8594 & 0.4767\\
& 120 &2.6499 & 0.4738 & 2.6498 & 0.4738 & 2.6495 & 0.4738 & 2.6499 & 0.4738\\
\hline
\end{tabular}}
\caption{Table of the generating and calibrated model prices and implied volatilities.}\label{table: compatibility options data}
\end{table}
\subsubsection{Implied Volatility and Monte Carlo Plots}
\begin{figure}[H]
\centering
	\subfigure[]{\includegraphics[width=0.45\textwidth]{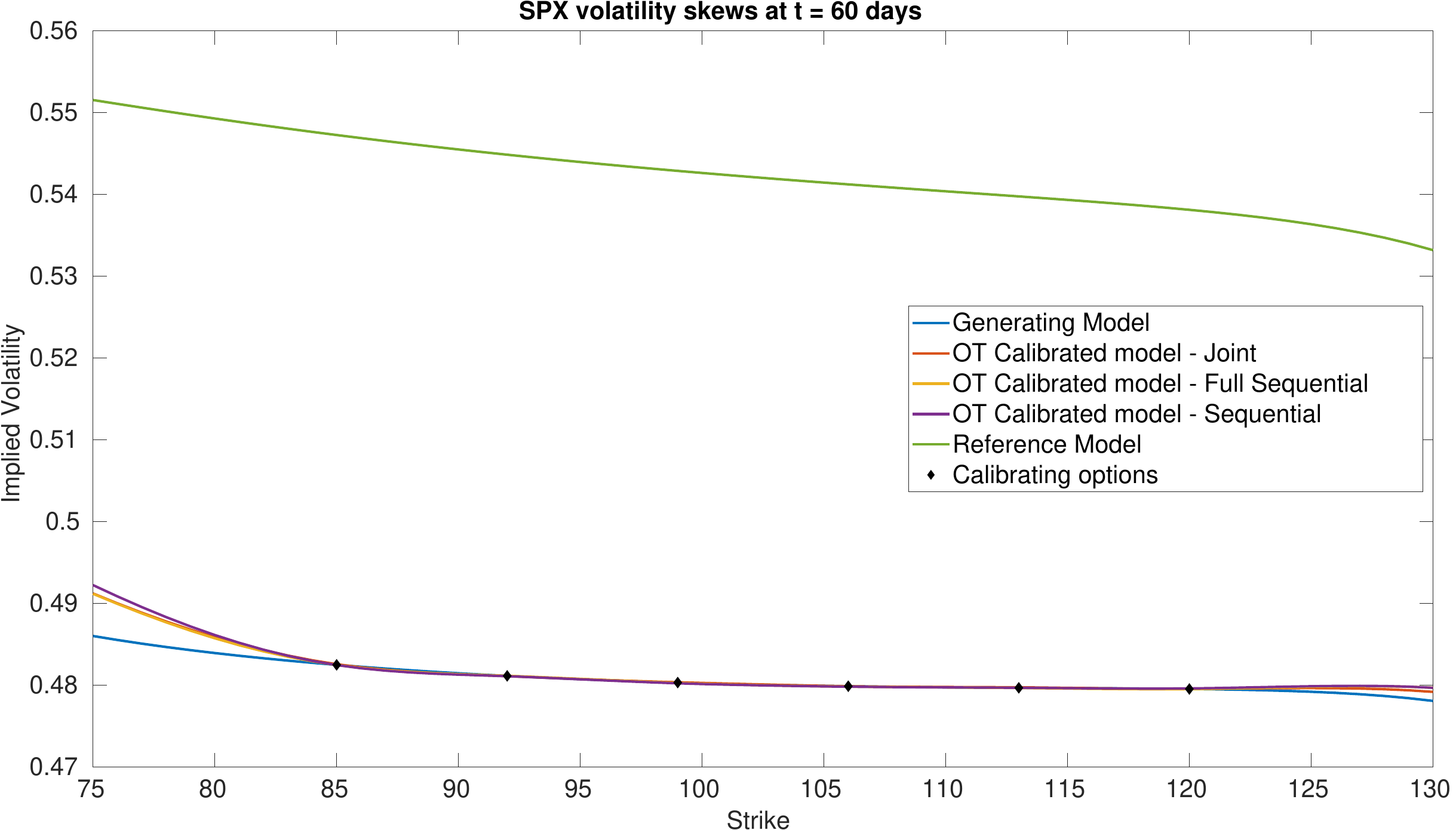} }
	\subfigure[]{\includegraphics[width=0.45\textwidth]{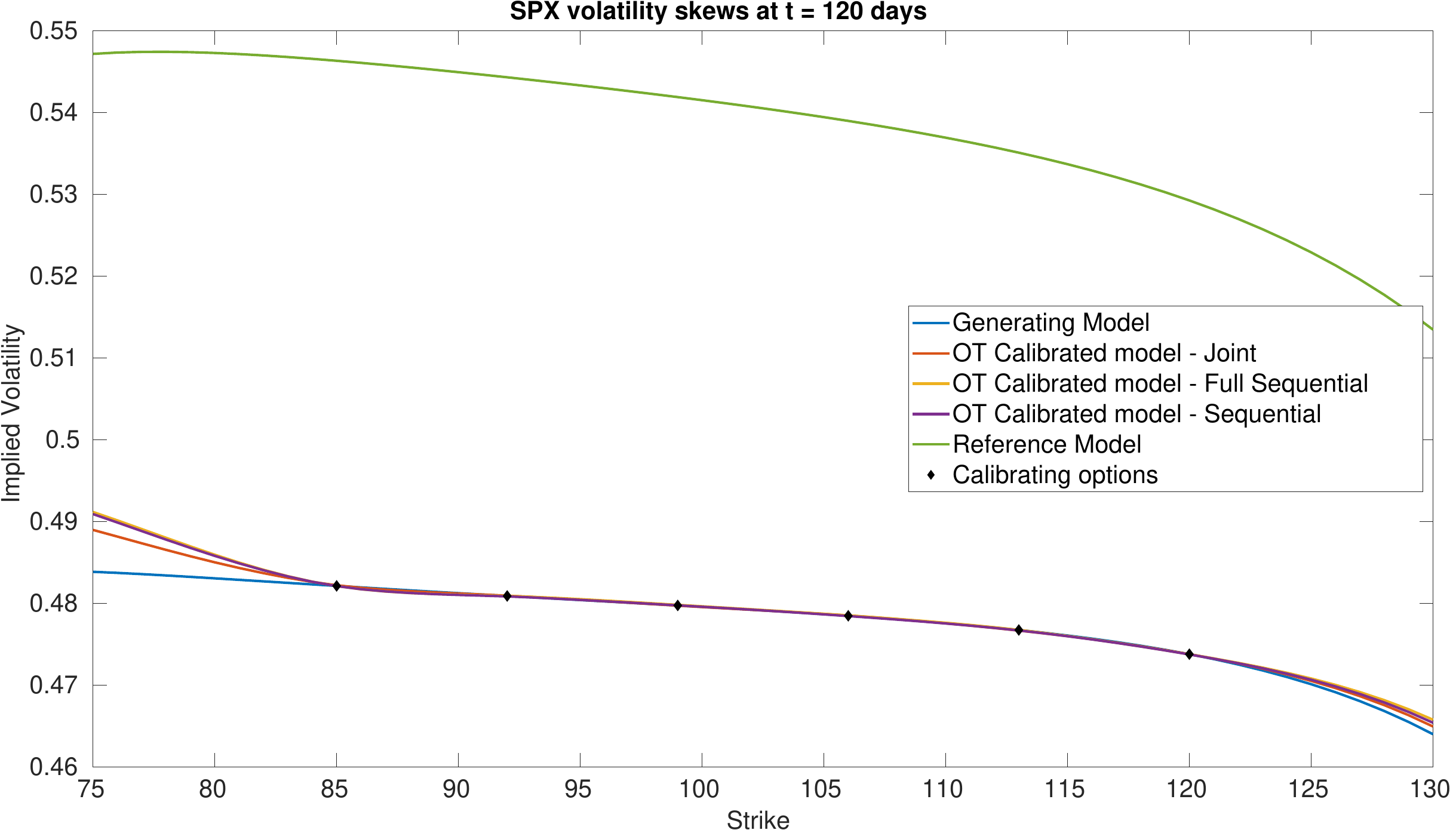} }
	\caption{Implied volatility skews of the SPX calibrating options under the reference model and calibrated HW-CEV model.}
	\label{fig: smile compatibility}
\end{figure}
\begin{figure}[H]
\centering
	\subfigure[]{\includegraphics[width=0.24\textwidth]{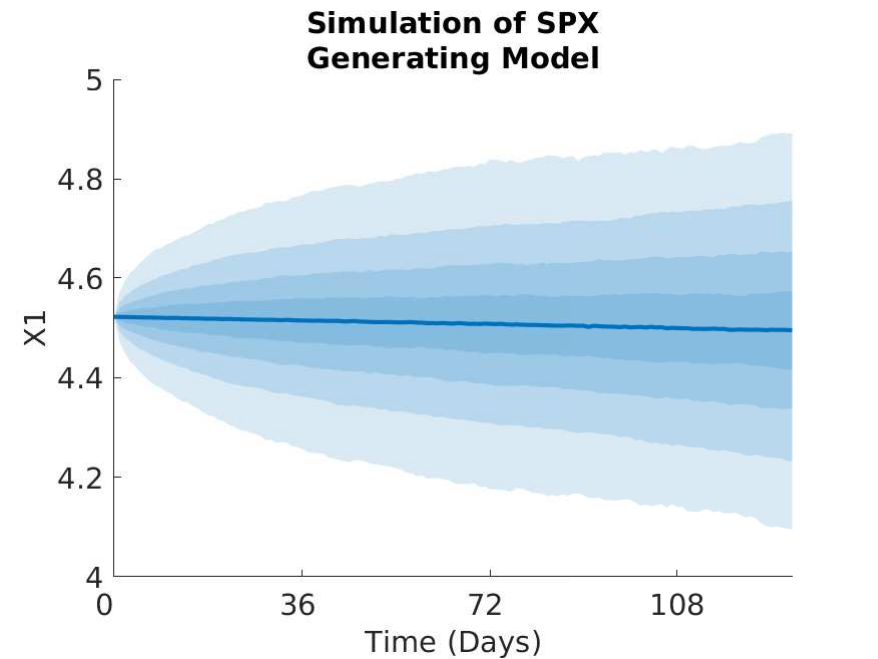}} 
	\subfigure[]{\includegraphics[width=0.24\textwidth]{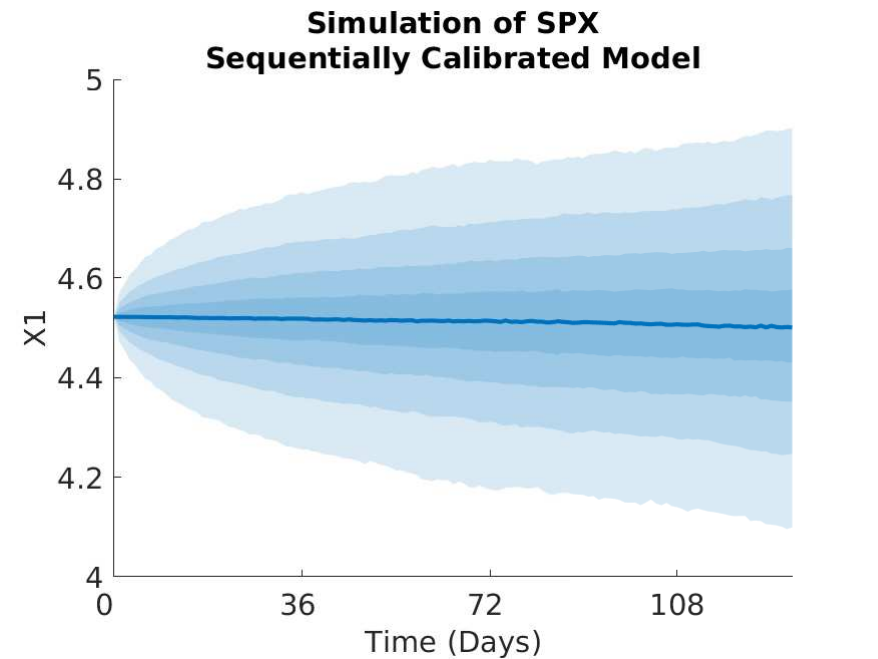}}
	\subfigure[]{\includegraphics[width=0.24\textwidth]{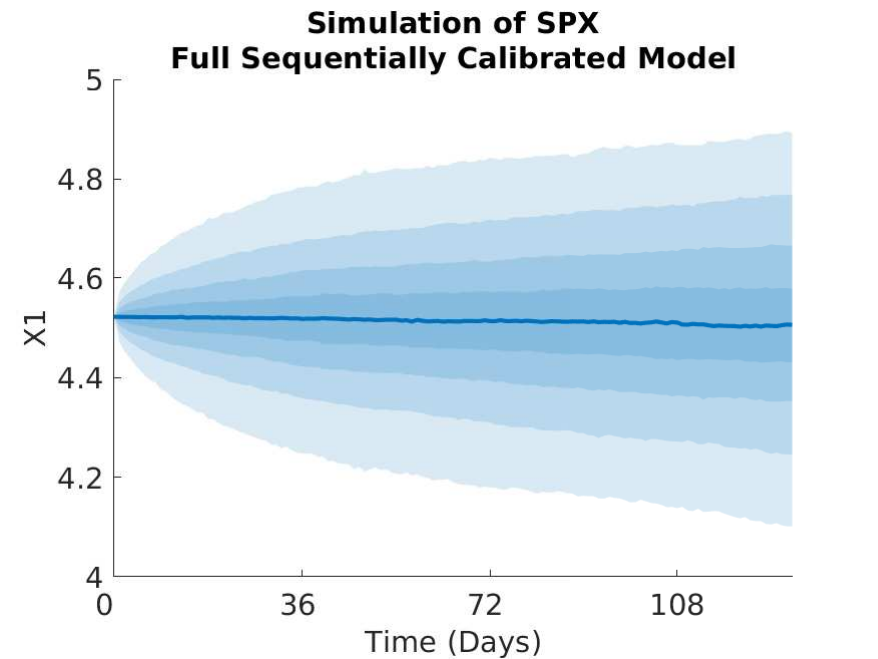}}
	\subfigure[]{\includegraphics[width=0.24\textwidth]{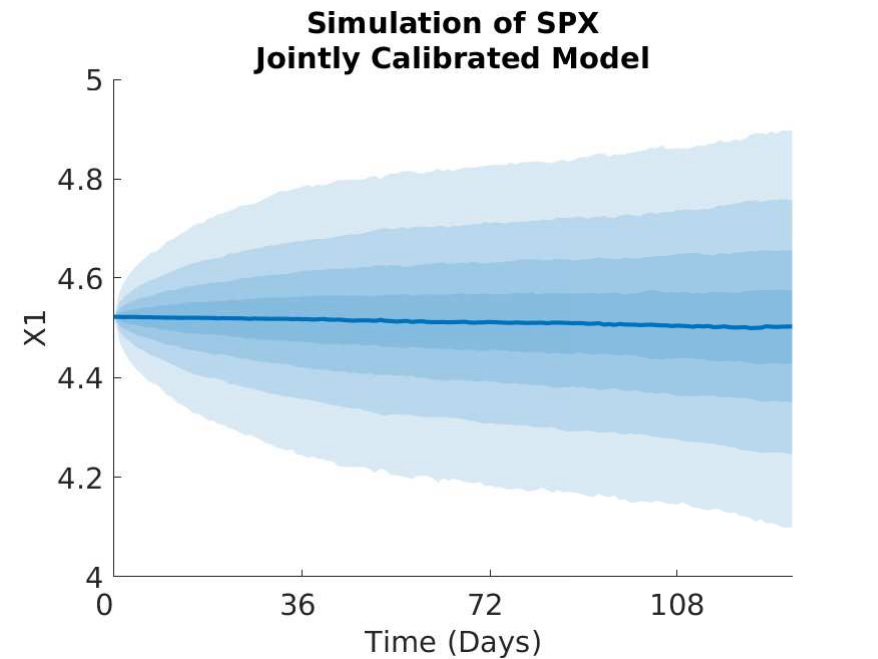}}

	\caption{Monte Carlo simulations of the SPX in the generating and calibrated HW-CEV models.}
\end{figure}
\subsubsection{Plots of the Volatility and Correlation surfaces}
For notational ease in the comparison, denote via a superscript $S,FS,J,G$ the characteristic under the SOT sequential calibrated model, SOT full sequential calibrated model, SOT jointly calibrated model and generating model respectively. In order to provide a meaningful comparison of the three methods, we show plots of the difference in the surfaces for all five characteristics, and how all of them differ from the generating model. Plots of the actual surfaces are left in the appendix. We omit all plots for $\alpha_1$ since in all three cases it is entirely specified by $\beta_{11}$. We remark that the formulae for $\beta_{11}^{FS}$ and $\beta_{11}^{J}$ given in Lemma~\ref{lem: optimisers joint} and Lemma~\ref{lem: optimal full sequential} are upon a first glance the same, and both differ from the formula for $\beta_{11}^{S}$ given in \eqref{eq: optimal beta11}. However, owing to the different cost functions and the fact that the joint problem allows for perturbations in the interest rate, whereas the full sequential problem does not, there is no expectation of the optimisers $\phi^*_{FS}$ and $\phi^*_{J}$ being the same since the global optimisers of both problems may differ. We expect the correlation coefficient $\rho^{S}$ to be almost constant and close to the reference model, owing to the assumption on the correlation in \eqref{eq: convexity adjustment}, whereas we expect some deviation away from the reference model for $\rho^{FS}$ and $\rho^{J}$ since we relax the assumptions on the correlation. The short rate volatility $\beta_{22}$ and short rate drift $\alpha_2$ are assumed to be a priori correct due to the matching generating and reference models. Note that the coefficients $\alpha_2^{S},\alpha_2^{FS},\beta_{22}^{S},\beta_{22}^{FS}$ are all fixed, and equal to these in the generating model, which we recall is also the reference model, i.e., we have\footnote{We confirmed this numerically in our results.} $\alpha_2^S=\alpha_2^{FS}=\alpha_2^G$ and $\beta_{22}^S=\beta_{22}^{FS}=\beta_{22}^G$. However, in the joint calibration case we do not fix $\alpha_2^J$ and $\beta_{22}^J$ and our optimisers given in Lemma~\ref{lem: optimisers joint} do not guarantee replication of the generating model. Nonetheless, since the cost function penalises deviations from the reference model, we expect that $\alpha_2^J$ and $\beta_{22}^J$ will be close to the generating model with some small perturbations away fro $\alpha_2^G$ and $\beta_{22}^G$.
\begin{figure}[H]
\centering
	\subfigure[]{\includegraphics[width=0.8\textwidth]{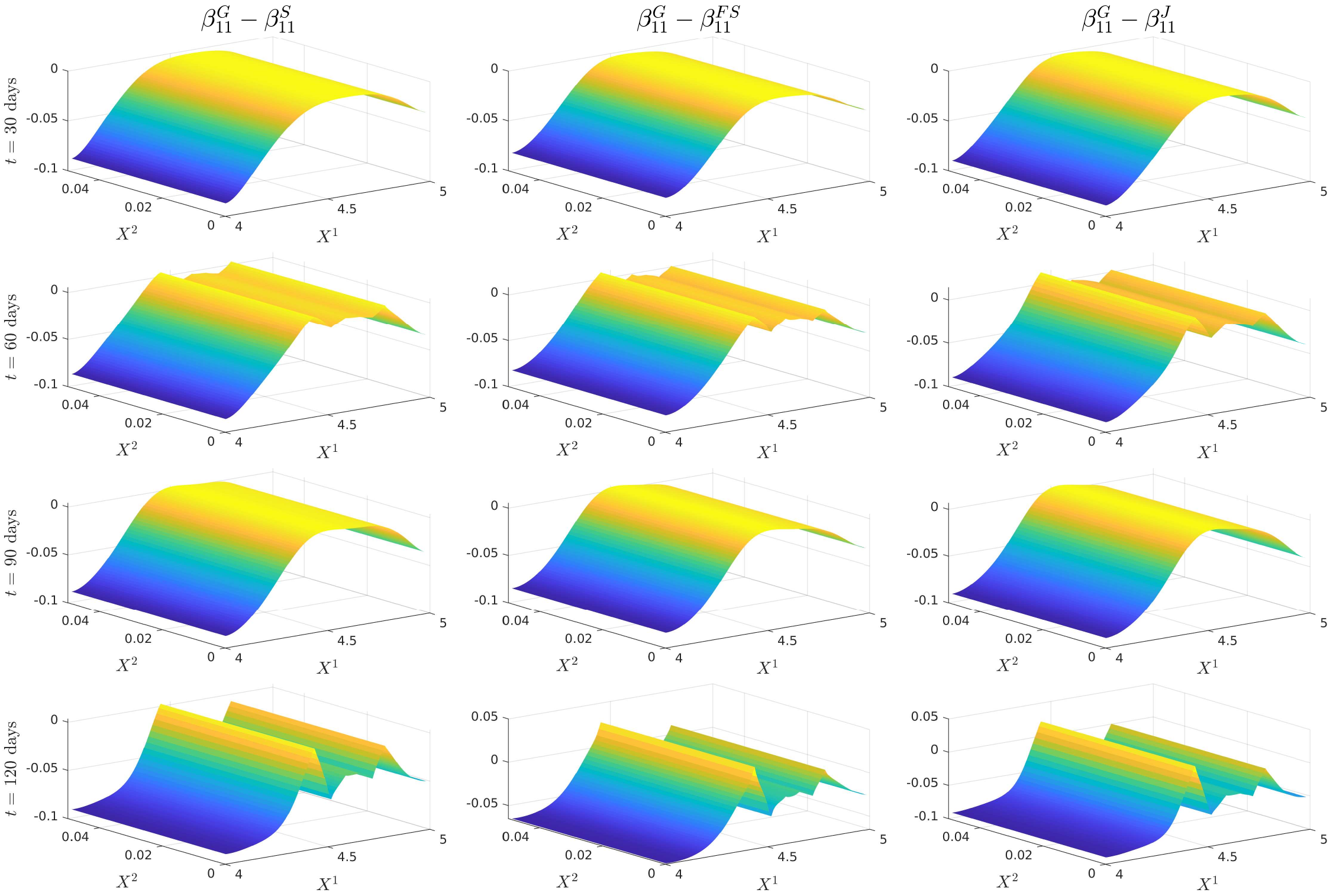}}
	\subfigure[]{\includegraphics[width=0.8\textwidth]{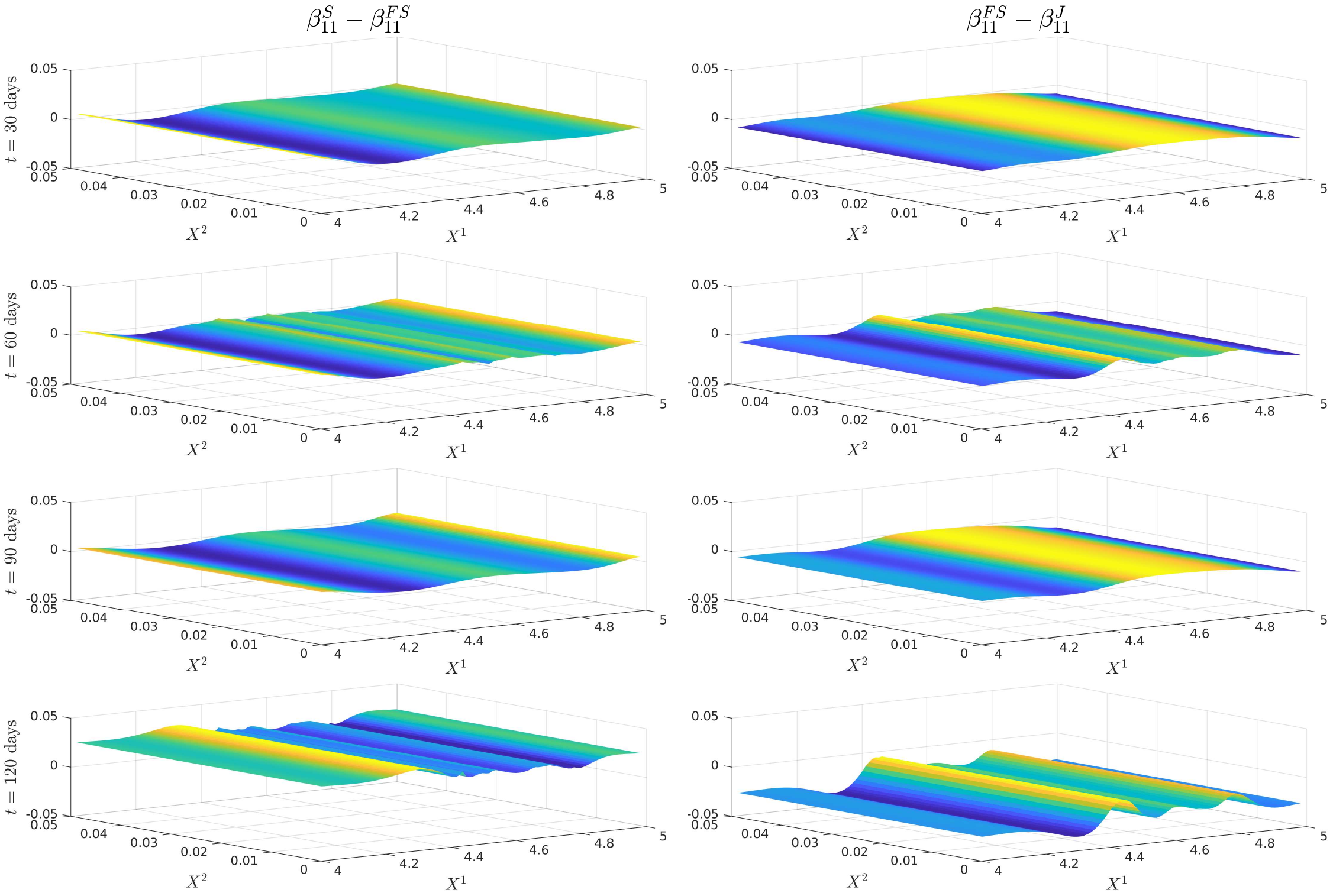}}
\caption{Plots of the difference between the generating model $\beta_{11}$ and SOT calibrated $\beta_{11}$ in the three methods (a), and plots of the difference between calibrated $\beta_{11}$ in the three methods (b).}
\label{fig: beta11 compatibility}
\end{figure}
\begin{figure}[H]
\centering
	\subfigure[]{\includegraphics[width=0.8\textwidth]{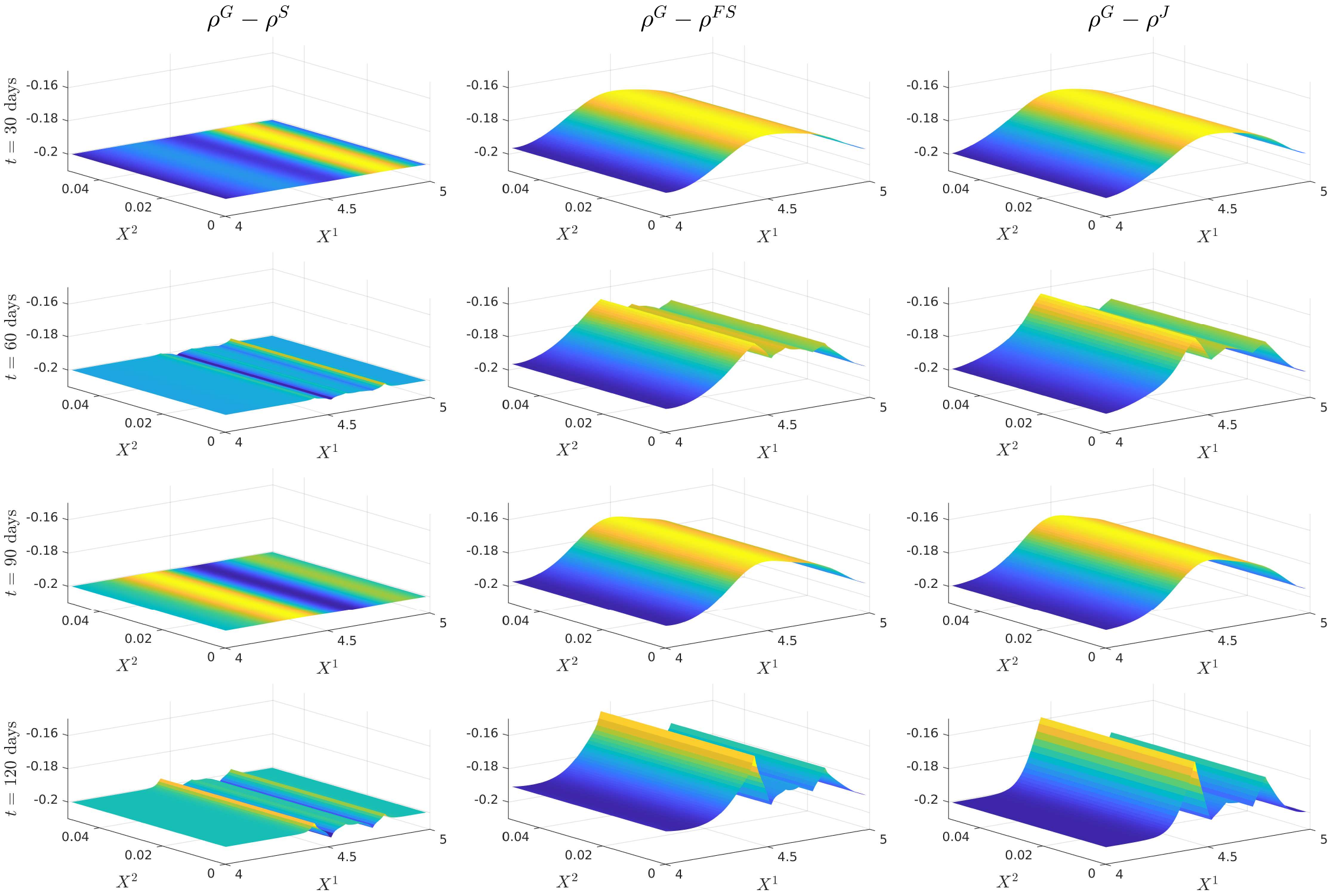}}
	\subfigure[]{\includegraphics[width=0.8\textwidth]{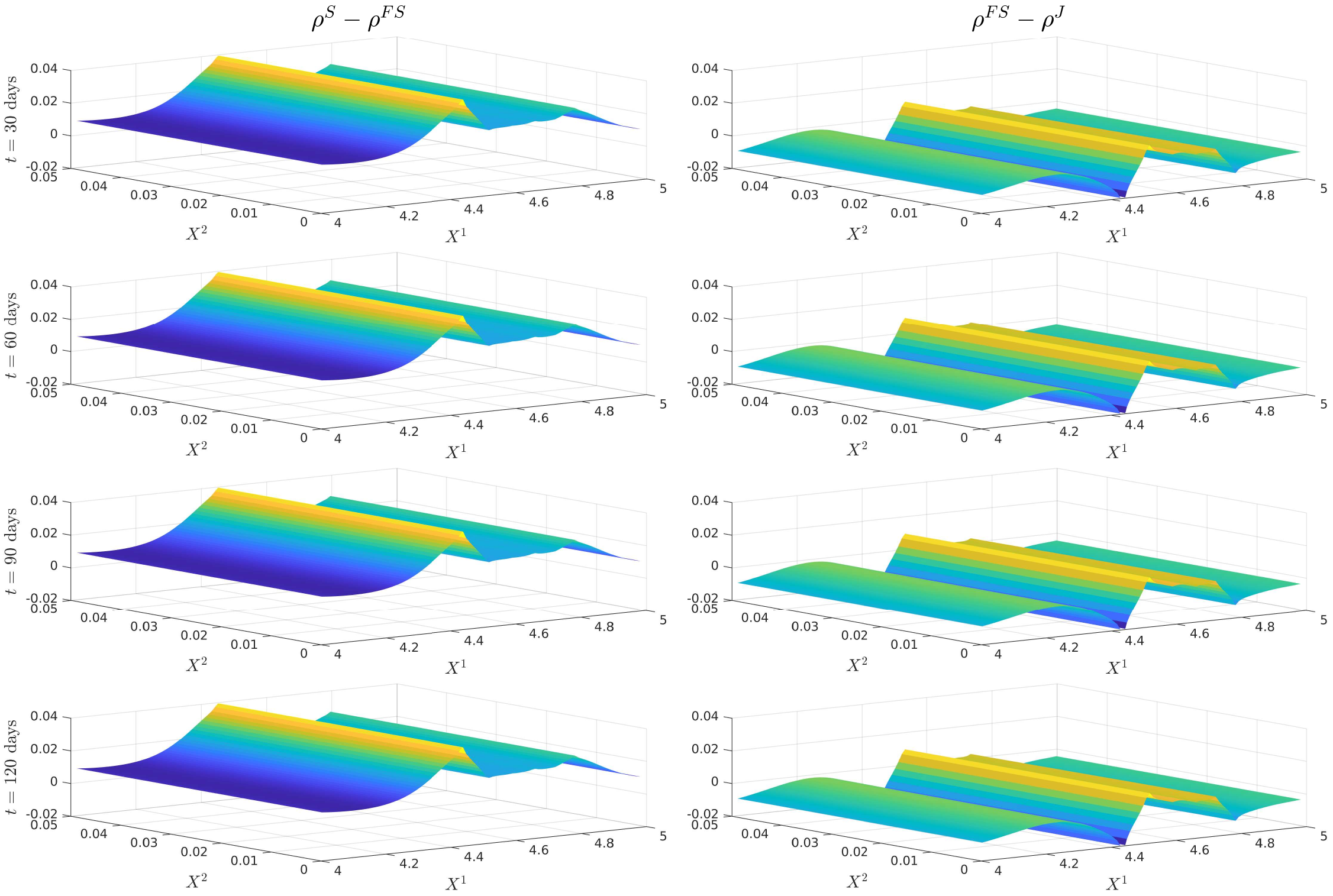}}
\caption{Plots of the difference between the generating model $\rho=\frac{\beta_{12}}{\sqrt{\beta_{11}\beta_{22}}}$ and SOT calibrated $\rho$ in the three methods (a), and plots of the difference between calibrated $\rho$ in the three methods (b). Note that we undid the scaling in $\beta_{12}$ and $\beta_{22}$ to recover the correlation between the log-stock and the short rate.}
\label{fig: rho compatibility}
\end{figure}
\begin{figure}[H]
\centering
\includegraphics[width=0.8\textwidth]{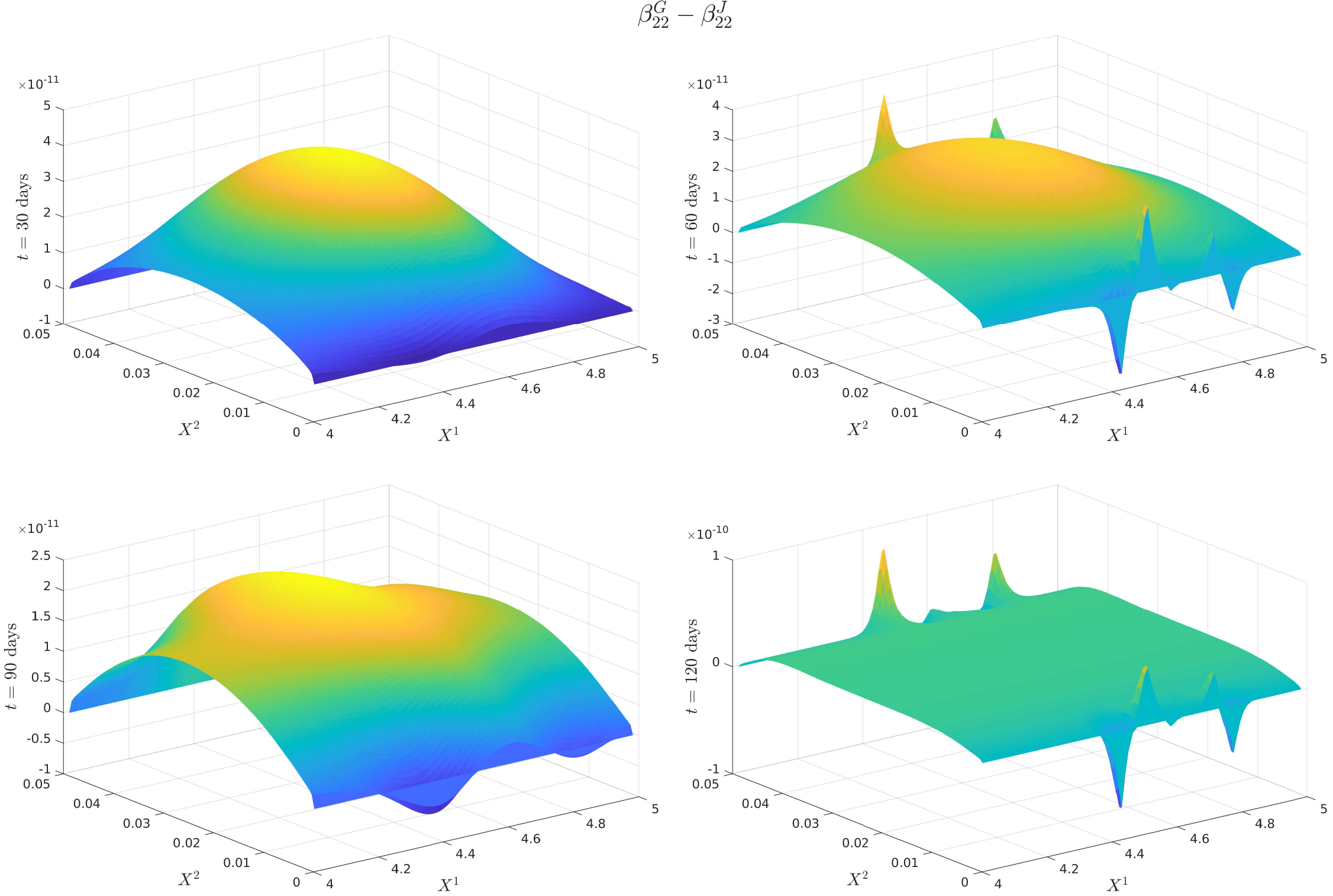}
\caption{Plots of the difference between the generating model $\beta_{22}$ and SOT jointly calibrated $\beta_{22}$. Note that we undid the scaling in $\beta_{22}$ to recover the volatility of the short rate.}
\label{fig: beta22 compatibility}
\end{figure}
\begin{figure}[H]
\centering
\includegraphics[width=0.8\textwidth]{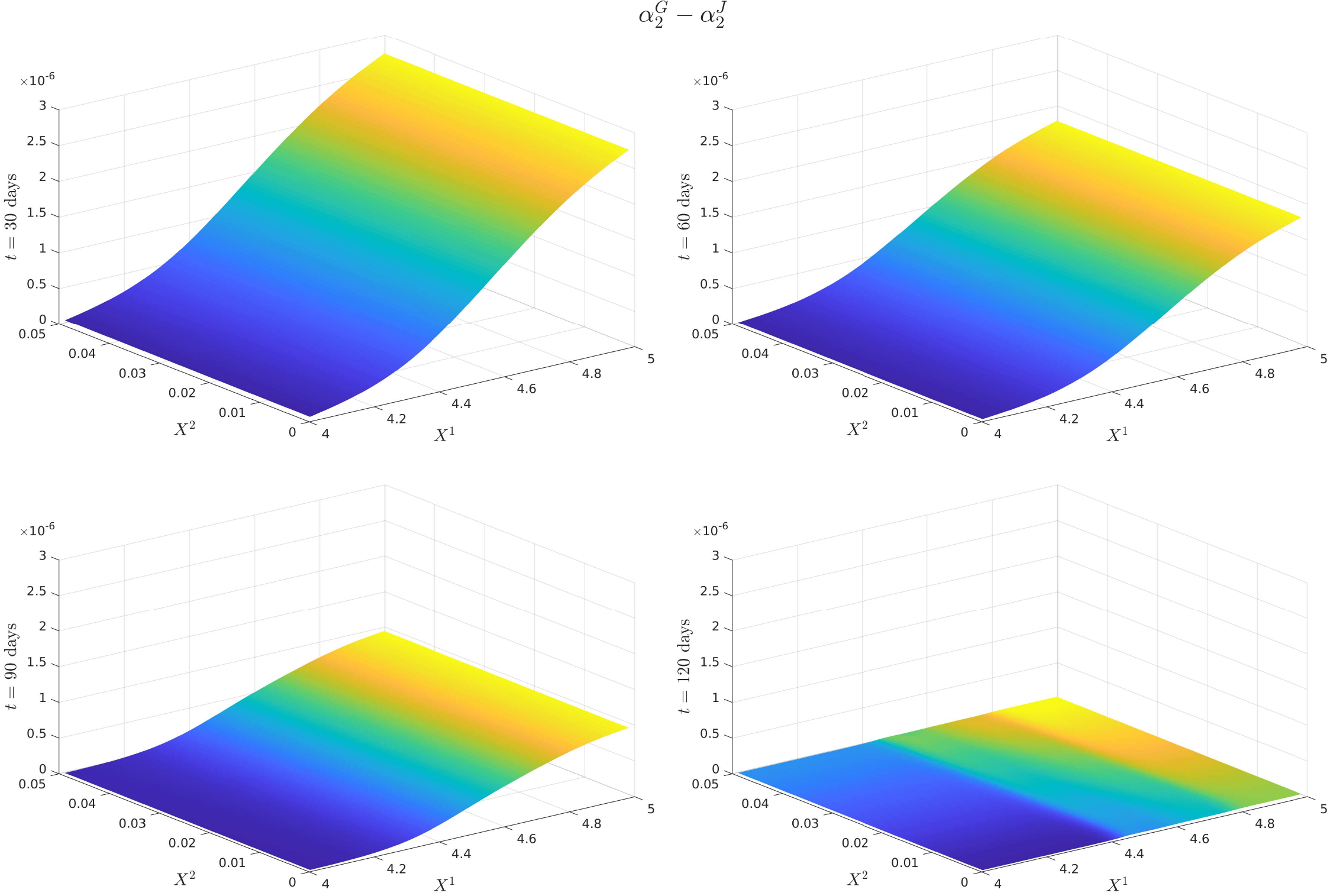}
\caption{Plots of the difference between the generating model $\alpha_{2}$ and SOT jointly calibrated $\alpha_{2}$.  Note that we undid the scaling in $\alpha_{2}$ to recover the drift of the short rate.}
\label{fig: alpha2 compatibility}
\end{figure}

\subsection{Discussion of the results}
Figure~\ref{fig: beta11 compatibility} (a) demonstrates that $\beta_{11}^S,\beta_{11}^{FS},\beta_{11}^J$ all are near the generating model within the region of our strikes and then deviate in the same manner towards the reference model outside the region of calibrating call option strikes. This is expected behaviour since in all three cases, the cost functions penalise deviations away from a reference model volatility while enforcing that the observed call option prices are replicated. A remarkable result is that Figure~\ref{fig: beta11 compatibility} (b) shows that $\beta_{11}^S,\beta_{11}^{FS},\beta_{11}^J$ are all close at $t=30,60,90$ days despite being defined differently. We do however see some expected perturbations on those dates and a larger perturbation at $t=120$ days which is also observed in the larger fluctuations seen in Figure~\ref{fig: beta11 compatibility} (a). Figure~\ref{fig: rho compatibility} (a) demonstrates strong dependence on the reference model: $\rho^S,\rho^{FS},\rho^J$ all differ from $\rho^{G}$ by approximately $-0.2$, however $\rho^{FS}$ and $\rho^J$ do perturb slightly towards to generating model in the region of strikes. Since our calibrating option has a payoff function only explicitly depending on the log-stock, this was expected behaviour since $\rho^{FS}$ and $\rho^{J}$ are defined via $\partial^2_{zr}\phi$. As remarked in \cite{JLO23calibration}, the extremely strong dependence on the reference model for $\rho^S$ arises mainly from the assumption on the form of the correlation given in (\ref{eq: convexity adjustment}). The differences observed in Figure~\ref{fig: rho compatibility} (b) arise from a fundamentally different handling of $\rho^S$ and that the volatility of the short rate is fixed so that $\rho^{FS}$ will be different to $\rho^J$. Nonetheless, since all three depend strongly on the reference model, these fluctuations are small. 

All three methods converged to a good accuracy with a maximum error in implied volatility of $10^{-4}$, as shown in Table~\ref{table: compatibility options data} and Figure~\ref{fig: smile compatibility}, where in all three cases, we applied Algorithm~\ref{algo: dual} with the smoothed reference model iteration. A minimum of 10 smoothing iterations were applied in all three cases to generate smoother surfaces. Since each smoothing iteration terminates either when calibration error is achieved or when a threshold of function evaluations were attained, in practice around 20 were required for the sequential calibration case to converge, as opposed to full sequential and joint calibration which both converged on the $10^{\mathrm{th}}$ smoothing iteration. The fastest in terms of computational time was the full sequential approach, taking just over an hour and the slowest was the joint calibration approach taking around three hours. Each epoch of a smoothed reference iteration was the fastest in sequential calibration. Joint calibration being the slowest was to be expected since it still computes the optimisers $\alpha^*_2$ and $\beta^*_{22}$ at each point, which is more expensive that simply fixing them as in the other two methods. 

\appendix
\section{Plots of Calibrated Characteristics from the Comparison}
We only show the plots of $\beta_{11}$ and $\rho$ since $\alpha_1$ is entirely determined by $\beta_{11}$ and Figure~\ref{fig: beta11 compatibility} and Figure~\ref{fig: alpha2 compatibility} demonstrate that all three cases are extremely close or exactly equal to the generating model for $\beta_{22}$ and $\alpha_2$.
\begin{figure}[H]
\centering
	\subfigure[]{\includegraphics[width=0.4\textwidth]{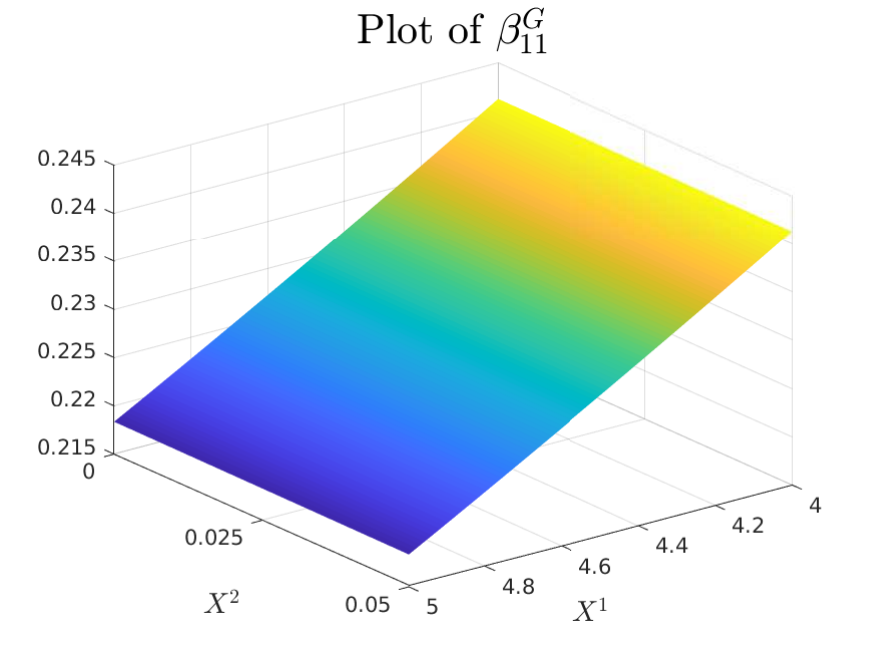}}
	\subfigure[]{\includegraphics[width=0.4\textwidth]{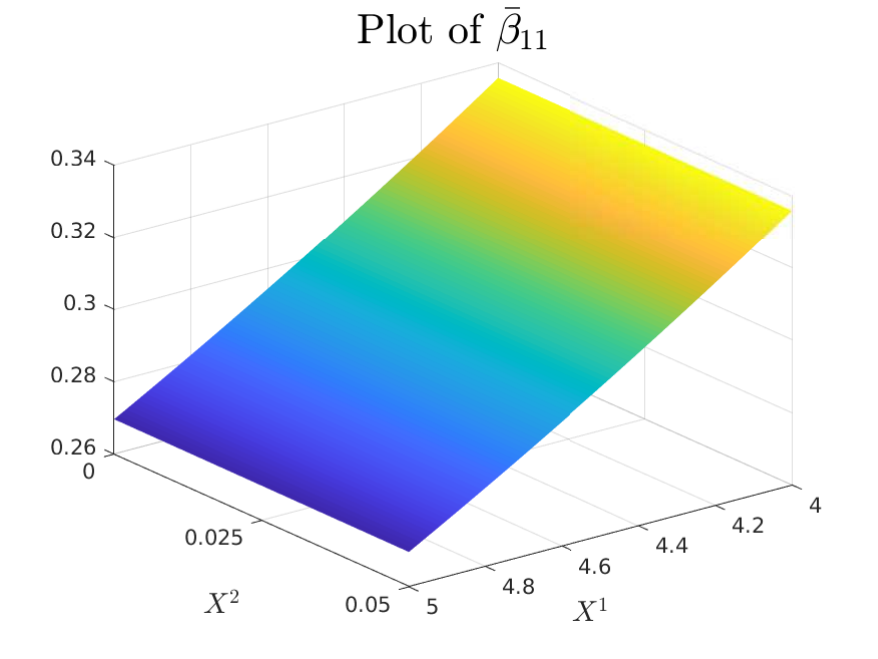}}
\caption{Plots of the generating model log-stock volatility $\beta_{11}^G$ (a), and reference model log-stock volatility $\overline{\beta}_{11}$ (b).}
\end{figure}
\begin{figure}[H]
\centering
\includegraphics[width = 0.8\textwidth]{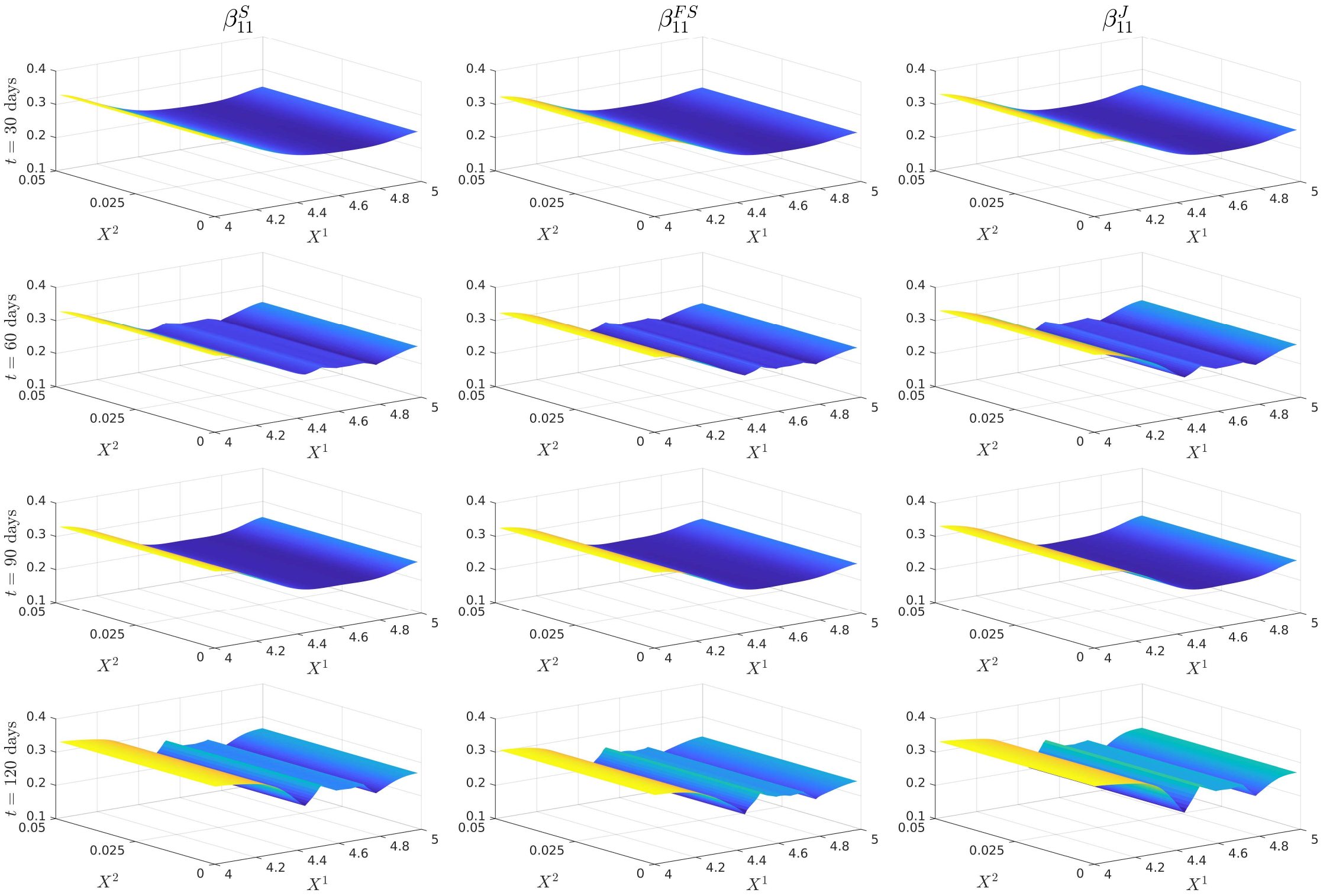}
\caption{Plots of calibrated $\beta_{11}$ in all three cases.}
\end{figure}
\begin{figure}[H]
\centering
\includegraphics[width = 0.8\textwidth]{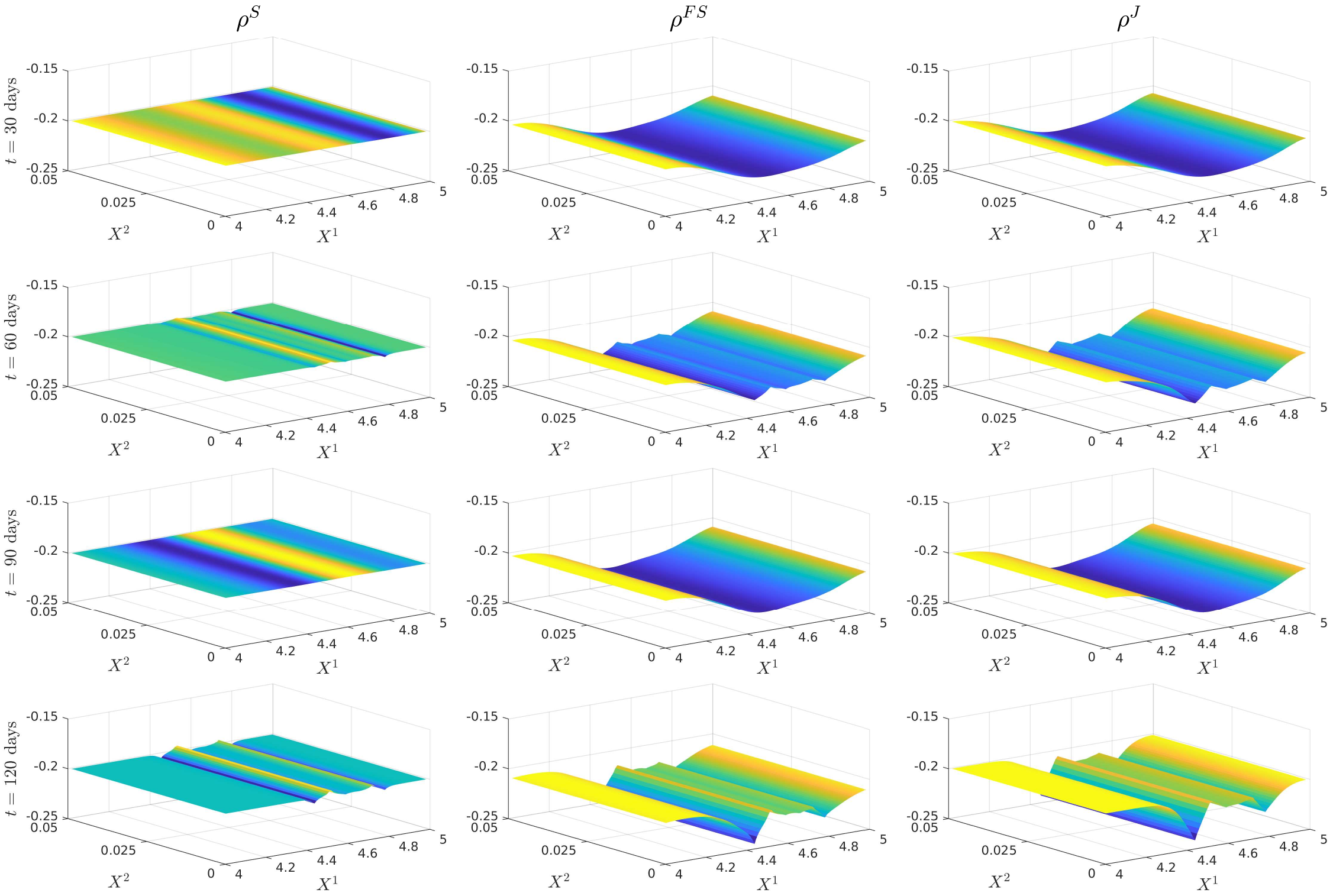}
\caption{Plots of calibrated $\rho=\frac{\beta_{12}}{\sqrt{\beta_{11}\beta_{22}}}$ in all three cases, where the scaling has been removed from $\beta_{12}$ and $\beta_{22}$. The generating model correlation $\rho^G=-0.4$ and the reference model correlation $\overline{\rho}=-0.2$.}
\end{figure}
\section{Proof of Lemma \ref{lem: optimisers joint}} \label{sec:appendix opt}
Since $\Gamma$ defined in (\ref{eq: gamma joint}) enforces that $\alpha^*_1 = r - \frac{1}{2}\beta^*_{11}$, we immediately have the first equality given an optimiser $\beta^*_{11}$. We remark that any reference model $(\overline{\alpha},\overline{\beta})$ will also follow the same constraint. Additionally, $\beta\in\IS^2_+$ is enforced by the set $\Gamma$. Therefore, since our cost function is given by (\ref{eq: cost function}), its Legendre-Fenchel transform $F^*:\IR^2\times\IS^2\to\IR$ is given by:
\begin{align}
F^*(a,b)&=\sup_{\substack{\alpha_2\in\IR,\\ \beta\in\IS^2_+,\\ \beta_{11}\in[\delta_{11}^l,\delta_{11}^u],\\ \beta_{22}\in[\delta_{22}^l,\delta_{22}^u]}}
\biggl\{a_1\left(r-\frac{1}{2}\beta_{11}\right)+a_2\alpha_2+b_{11}\beta_{11}+2b_{12}\beta_{12}+b_{22}\beta_{22}-\frac{1}{4}(\beta_{11}-\overline{\beta}_{11})^2-(\alpha_2-\overline{\alpha}_2)^2\notag\\
&\qquad\qquad\qquad\qquad-(\beta_{11}-\overline{\beta}_{11})^2-2(\beta_{12}-\overline{\beta}_{12})^2-(\beta_{22}-\overline{\beta}_{22})^2\biggr\}\label{eq: LF cost fct}
\end{align}
The gradient $\nabla F^*(a,b)$ is given by the maximisers of (\ref{eq: LF cost fct}), and by rearranging the $\alpha_2$ term, we obtain the second equality:
$$
\alpha_2^* = \argmax_{\alpha_2\in\IR}\left\{a_2\alpha_2-\alpha_2^2+2\alpha_2\overline{\alpha}_2-\overline{\alpha}^2_2\right\}=\argmin_{\alpha_2\in\IR}\left\{\left(\alpha_2-\left(\overline{\alpha}_2+\frac{1}{2}a_2\right)\right)^2\right\}=\overline{\alpha}_2+\frac{1}{2}a_2.
$$
Now, similarly rearranging (\ref{eq: LF cost fct})
\begin{align*}
[\beta_{12}^*,\beta_{11}^*,\beta_{22}^*]&=\argmin_{\substack{\beta_{12}\\ \beta\in\IS^2_+}}\Biggl\{\Biggl[\left(\beta_{12}-\left(\overline{\beta}_{12}+\frac{1}{2}b_{12}\right)\right)^2,\argmin_{\substack{\beta_{11}\\ \beta_{11}\in[\delta_{11}^l,\delta_{11}^u]}}\left\{\left(\beta_{11}-\left(\overline{\beta}_{11}+\frac{1}{5}\left(2b_{11}-a_1\right)\right)\right)^2\right\},\\
&\qquad\qquad\qquad\qquad\argmin_{\substack{\beta_{22}\\\beta_{22}\in[\delta_{22}^l,\delta_{22}^u]}}\left\{\left(\beta_{22}-\left(\overline{\beta}_{22}+\frac{1}{2}b_{22}\right)\right)^2\right\}\Biggr]\Biggr\}
\end{align*}
We solve the minimisations of $\beta_{11}$ and $\beta_{22}$ inside the $\beta_{12}$ minimisation problem by taking:
\begin{align*}
\beta^*_{11} &= \begin{cases}\overline{\beta}_{11}+\frac{1}{5}(2b_{11}-a_1),&\text{ if } \overline{\beta}_{11}+\frac{1}{5}(2b_{11}-a_1)\in [\delta_{11}^l,\delta_{11}^u]\\
\delta_{11}^u,&\text{ if }\overline{\beta}_{11}+\frac{1}{5}(2b_{11}-a_1)>\delta_{11}^u\\
\delta_{11}^l,&\text{ if }\overline{\beta}_{11}+\frac{1}{5}(2b_{11}-a_1)<\delta_{11}^l\end{cases}\\
\beta^*_{22}&=\begin{cases}\overline{\beta}_{22}+\frac{1}{2}b_{22},&\text{ if }\overline{\beta}_{22} + \frac{1}{2}b_{22}\in[\delta_{22}^l,\delta_{22}^u],\\
\delta_{22}^u,&\text{ if }\overline{\beta}_{22} + \frac{1}{2}b_{22} >\delta_{22}^u,\\
\delta_{22}^l,&\text{ if }\overline{\beta}_{22} + \frac{1}{2}b_{22} < \delta_{22}^l.\end{cases} 
\end{align*}
Given $\beta^*_{11}$ and $\beta^*_{22}$ as above, we now may rewrite the outside constraint as $\beta_{12}\in[-\sqrt{\beta^*_{11}\beta^*_{22}},\sqrt{\beta^*_{11},\beta^*_{22}}]$ and thus obtain:
$$
\beta^*_{12}=\begin{cases}\overline{\beta}_{12}+\frac{1}{2}b_{12},&\text{ if }\overline{\beta}_{12}+\frac{1}{2}b_{12}\in[-\sqrt{\beta^*_{11}\beta^*_{22}},\sqrt{\beta^*_{11}\beta^*_{22}}],\\
\sqrt{\beta^*_{11}\beta^*_{22}},&\text{ if } \overline{\beta}_{12}+\frac{1}{2}b_{12}>\sqrt{\beta^*_{11}\beta^*_{22}},\\
-\sqrt{\beta^*_{11}\beta^*_{22}},&\text{ if } \overline{\beta}_{12}+\frac{1}{2}b_{12} < -\sqrt{\beta^*_{11}\beta^*_{22}}.\end{cases}
$$
In particular, in the first case, the condition $\beta\in \IS^2_+$ is automatically satisfied and the procedure returns the optimizer. 
By taking $a=\nabla_x\phi$ and $b=\frac{1}{2}\nabla_x^2\phi$, we conclude the proof.
\section{Application of Full Sequential Calibration to Local-Stochastic Volatility Models}
We observe there that the method of Section \ref{sect: fs calib} can be applied to the local-stochatic volatility calibration setting of \cite{guo2022calibration} to relax the constraint on the correlation imposed in that paper. In this setting, we assume our interest rate is zero and that our reference model is given by a Heston model, derived in \cite{heston1993closed}. The state variables are therefore our log-stock and a correlated stochastic term for the volatility, with dynamics given by:
\begin{align*}
\diff Z_t&= -\frac{1}{2}V_t\diff t + \sqrt{V_t}\diff W_t^1,\\
\diff V_t&= \kappa (\theta - V_t)\diff t + \xi\sqrt{V_t}\diff W^2_t,\\
\diff\langle W^1,W^2\rangle_t&=\rho\diff t,
\end{align*}
where $\kappa,\theta,\xi>0$ and $\rho\in [-1,1]$. We assume that the state variable $V_t$ is known, so the model characteristics that we want to calibrate are the volatility of the log-stock and the correlation between the state variables. As before, the instruments that we calibrate this model to are European call options on the chosen stock. We now define our cost function in a similar manner to Section \ref{sect: fs calib} by defining the convex set $\Gamma_{\mathrm{LSV}}$:
\begin{equation}\label{eq: gamma LSV}
\Gamma_{\text{LSV}}(v)=\left\{(\alpha,\beta)\in\IR^2\times\IS^2_+\: :\:\alpha_1=-\frac{1}{2}\beta_{11},\:\alpha_2=\kappa(\theta - v),\:\beta_{22}=\xi^2v,\:\delta_{11}^l\leq\beta_{11}\leq\delta_{11}^u \right\}.
\end{equation} 
\begin{equation}\label{eq: LSV cost function}
F_{\text{LSV}}(v,\alpha,\beta)=\begin{cases}\frac{5}{4}(\beta_{11}-\overline{\beta}_{11})^2+2(\beta_{12}-\overline{\beta}_{12})^2,&\text{if }(\alpha,\beta)\in\Gamma_{\text{LSV}}(v),\\
+\infty,&\text{otherwise.}\end{cases}
\end{equation}
Since we have no stochastic discount factor, we no longer need the sub-probability measure approach of \cite{JLO23calibration}, and therefore can directly apply the duality result of \cite[Proposition~3.7]{guo2022calibration}. With our cost function defined in (\ref{eq: LSV cost function}), we therefore obtain the following dual formulation:
\begin{problem}[Full Sequential Dual Formulation]\label{prob: LSV dual}
$$
V=\sup_{\lambda}\lambda\cdot u-\phi^{\lambda}(0,Z_0,V_0).
$$
Where $\phi^{\lambda} = \phi(t,z,v)$ solves the HJB equation:
\begin{align}
\sum_{i=1}^n\lambda_iG_i(x)\delta_{\tau_i}+\partial_t\phi+&\sup_{\substack{\beta_{11},\beta_{12},\\ \beta\in\IS^2_+}}\biggl(-\frac{1}{2}\beta_{11}\partial_{z}\phi+\kappa(\theta-v)\partial_{v}\phi+\frac{1}{2}\beta_{11}\partial^2_{zz}\phi+\beta_{12}\partial^2_{zv}\phi+\frac{1}{2}\xi^2v\partial^2_{vv}\phi\notag\\&\qquad\qquad-F_{\mathrm{LSV}}(v,\alpha,\beta)\biggr)=0,\quad (t,z,v)\in [0,T]\times\IR^2.\label{eq: HJB LSV Full seq cal}
\end{align}
\end{problem}
Similar to Lemma~\ref{lem: optimal full sequential}, we obtain the following approximation for $\beta^*_{11}$ and $\beta^*_{12}$:
\begin{lemma}
Let $\beta^*_{22}= \xi^2 v$ and define $\beta_{11}^*$ and $\beta_{12}^*=\beta^*_{21}$ by 
\begin{align*}
\beta_{11}^*(t,z,v)&=\begin{cases}\overline{\beta}_{11}(t,z,v) + \frac{1}{5}(\partial^2_{zz}\phi(t,z,v)-\partial_z\phi(t,z,v)),&\text{ when } \overline{\beta}_{11} + \frac{1}{5}(\partial^2_{zz}\phi-\partial_z\phi)\in[\delta_{11}^l,\delta_{11}^u]\\
\delta_{11}^l,&\text{ when }\overline{\beta}_{11} + \frac{1}{5}(\partial^2_{zz}\phi-\partial_z\phi)<\delta_{11}^l,\\
\delta_{11}^u,&\text{ when }\overline{\beta}_{11} + \frac{1}{5}(\partial^2_{zz}\phi-\partial_z\phi)>\delta_{11}^u,\end{cases}\\
\beta_{12}^*(t,z,v)&=\begin{cases}\overline{\beta}_{12}(t,z,v)+ \frac{1}{4}\partial^2_{zv}\phi(t,z,v),&\text{ when } \overline{\beta}_{12} + \frac{1}{4}\partial^2_{zv}\phi\in\left[-\xi\sqrt{v\beta^*_{11}},\xi\sqrt{v\beta^*_{11}}\right],\\
-\xi\sqrt{v\beta^*_{11}},&\text{ when }\overline{\beta}_{12} + \frac{1}{4}\partial^2_{zv}\phi< -\xi\sqrt{v\beta^*_{11}},\\
\xi\sqrt{v\beta^*_{11}},&\text{ when }\overline{\beta}_{12} + \frac{1}{4}\partial^2_{zv}\phi>\xi\sqrt{v\beta^*_{11}}.\end{cases}
\end{align*}
Then $\beta^*$ is a positive semi-definite matrix and whenever 
$$
-\xi\sqrt{v\beta^*_{11}} < \beta^*_{12} < \xi\sqrt{v\beta^*_{11}}
$$
then $\beta^*$ is the optimizer in \eqref{eq: HJB LSV Full seq cal}. 
\end{lemma}

We numerically solve this problem using the same methods as in Section \ref{sec: numerics}, we show a table of parameters below. We simulate the call option prices from the generating model and test it against two reference models: a ``good'' reference model and a ``bad'' reference model. The parameters used in $\Gamma_{\mathrm{LSV}}$ are those of the reference model.
\begin{table}[H]
\centering
\begin{tabular}{|l|l|p{0.7\linewidth}|}
\hline
\multicolumn{3}{|l|}{Heston Model}\\
\hline
Parameter & Value & Interpretation\\
\hline
$X^1_0$ & $\log(92)$ & Initial log-stock price\\
$X^2_0$ & $0.25$ & Initial volatility\\
$\epsilon_1$ & $1\times 10^{-4}$ & Tolerance for the difference in scaled model and market implied volatility\\
$\epsilon_2$ & $1\times 10^{-12}$ & Tolerance for the policy iteration approximation of the optimal characteristics\\
\hline
$\kappa$ & 1 & Speed of volatility mean reversion in the generating model\\
$\theta$ & 0.05 & Long-term mean of the volatility in the generating model\\
$\xi$ & 0.2 & Volatility scaling of the volatility in the generating model\\
$\rho$ & -0.4 & Instantaneous correlation between the log-stock and volatility in the generating model\\
\hline
$\overline{\kappa}_{\mathrm{good}}$ & 1.5 & Speed of volatility mean reversion in the reference model\\
$\overline{\theta}_{\mathrm{good}}$ & 0.07 & Long-term mean of the volatility in the reference model\\
$\overline{\xi}_{\mathrm{good}}$ & 0.15 & Volatility scaling of the volatility in the reference model\\
$\overline{\rho}_{\mathrm{good}}$ & -0.2 & Instantaneous correlation between the log-stock and volatility in the reference model\\
\hline
$\overline{\kappa}_{\mathrm{bad}}$ & 2 & Speed of volatility mean reversion in the reference model\\
$\overline{\theta}_{\mathrm{bad}}$ & 0.09 & Long-term mean of the volatility in the reference model\\
$\overline{\xi}_{\mathrm{bad}}$ & 0.3 & Volatility scaling of the volatility in the reference model\\
$\overline{\rho}_{\mathrm{bad}}$ & 0.2 & Instantaneous correlation between the log-stock and volatility in the reference model\\
\hline
\end{tabular}
\caption{Parameter values of generating and reference models used in the LSV calibration.}
\end{table}
\begin{table}[h]
\centering
\begin{tabular}{|c|c|c|c|c|c|c|c|}
\hline
 & &\multicolumn{2}{c|}{Generating Model} & \multicolumn{2}{c|}{Calibrated Model:} & \multicolumn{2}{c|}{Calibrated Model:}\\
& & \multicolumn{2}{c|}{} &\multicolumn{2}{c|}{Good Reference} & \multicolumn{2}{c|}{Bad Reference}\\
\hline
Option Type&Strike&Price&IV&Price&IV&Price&IV\\
\hline
\multirow{6}{*}{\begin{tabular}{c}SPX Call options\\ $t=60$ days\end{tabular}} & 85 & 11.0144 & 0.4840 & 11.0143 & 0.4840 & 11.0148 & 0.4841\\
& 92 & 7.1928 & 0.4808 & 7.1928 & 0.4808 & 7.1920 & 0.4808\\
& 99 & 4.4416 & 0.4782 & 4.4419 & 0.4782 & 4.4414 & 0.4782\\
& 106 & 2.6037 & 0.4761 & 2.6036 & 0.4761 & 2.6034 & 0.4760\\
& 113 & 1.4564 & 0.4744 & 1.4562 & 0.4743 & 1.4562 & 0.4743\\
& 120 & 0.7813 & 0.4729 & 0.7814 & 0.4730 & 0.7811 & 0.4729\\
\hline
\multirow{6}{*}{\begin{tabular}{c}SPX Call options\\ $t=120$ days\end{tabular}} & 85 & 13.4256 & 0.4686 & 13.4260 & 0.4687 & 13.4256 & 0.4686\\
&92 & 9.8367 & 0.4656& 9.8372 & 0.4656 & 9.8370 & 0.4656\\
&99 & 7.0267 & 0.4628 & 7.0273 & 0.4628 & 7.0250 & 0.4627\\
&106 & 4.9025 & 0.4601 & 4.9043 & 0.4602 & 4.9034 & 0.4602\\
&113 & 3.3437 & 0.4574 & 3.3450 & 0.4575 & 3.3451 & 0.4575\\
&120 & 2.2243 & 0.4541 & 2.2249 & 0.4541 & 2.2252 & 0.4541\\
\hline
\end{tabular}
\caption{Table of the generating and calibrated model prices and implied volatilities.}\label{table: LSV options data}
\end{table}
\begin{figure}[H]
\centering
	\subfigure[]{\includegraphics[width=0.45\textwidth]{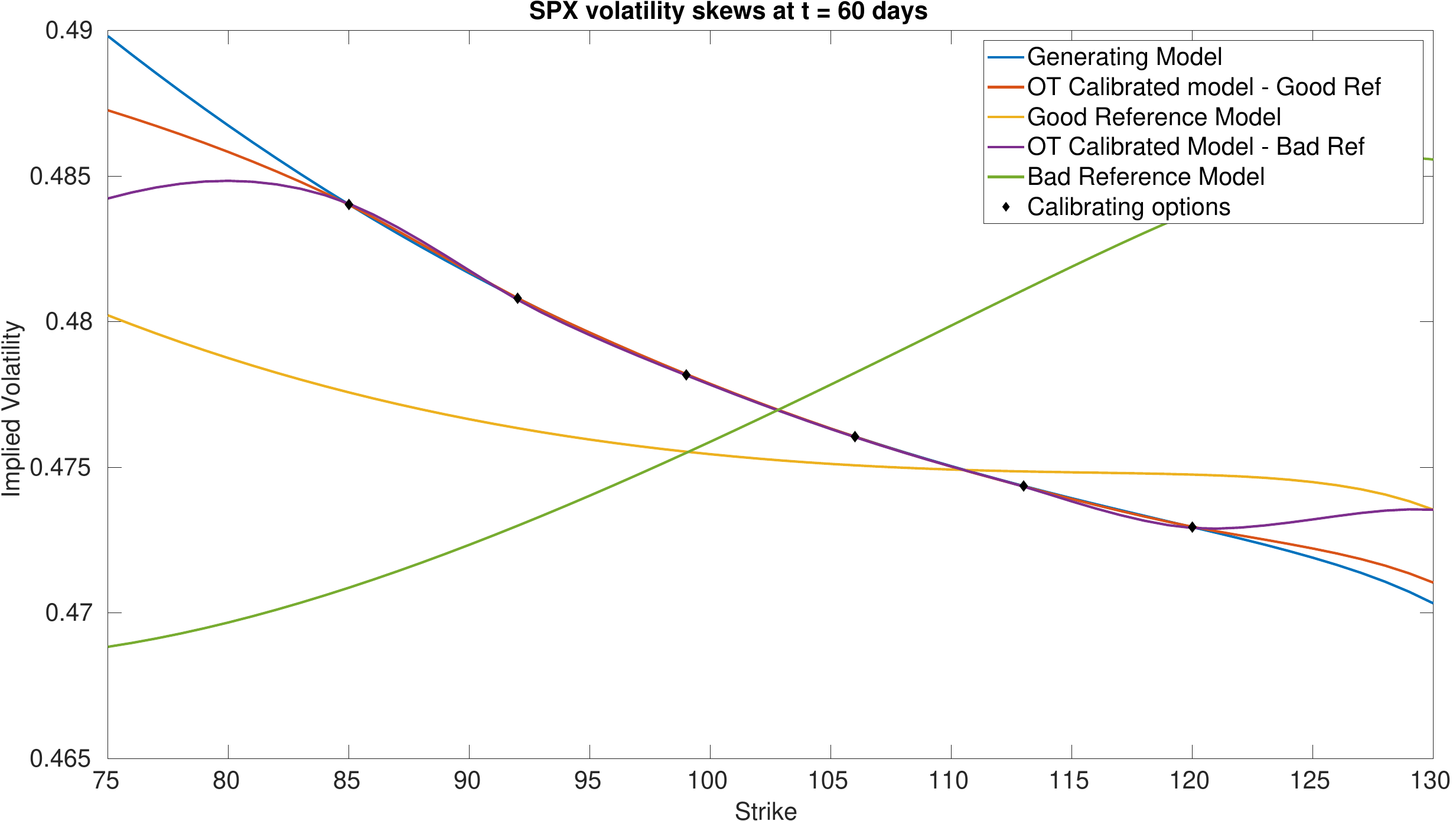} }
	\subfigure[]{\includegraphics[width=0.45\textwidth]{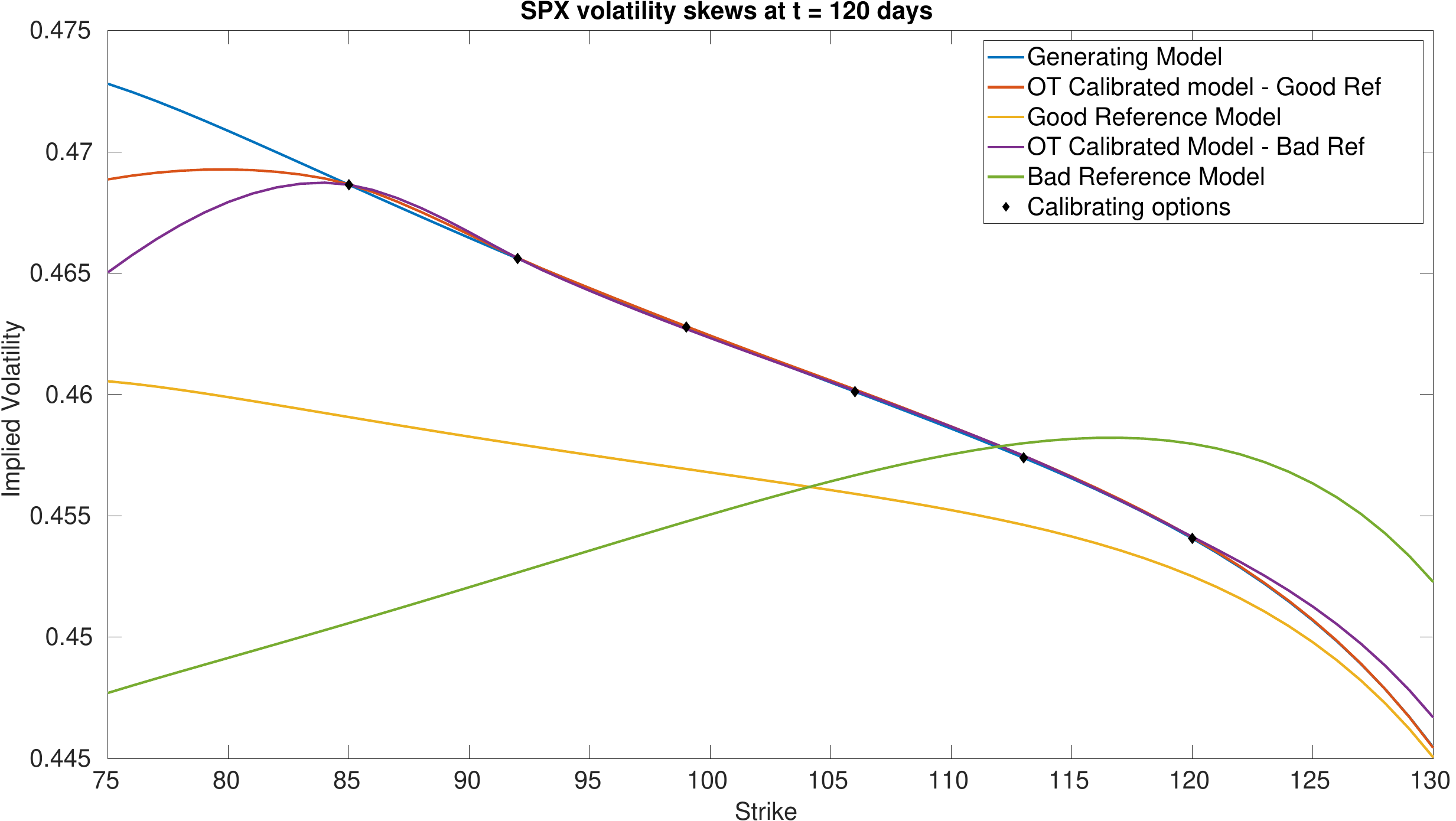} }
	\caption{Implied volatility plots for the generating model, both SOT calibrated models and both reference models.}
\end{figure}
We remark that this approach can recover the implied volatility within the range of the strikes even when the reference model has completely the wrong shape of implied volatility. We now display the surfaces of SOT calibrated $\beta_{11}$ and $\rho$. We notice that while the generating and reference models in $\beta_{11}$ are the same, our calibrated model is perturbed from the reference. We also notice that our calibrated correlation $\rho$ is close to the reference model correlation parameter in both cases, with deviations towards the generating model when we are in the range of our strikes and when $V_t$ is near zero. This strong dependence is also seen in our previous calibration approaches, and thus a good a priori estimate of $\rho$ would be needed in practice.
\begin{figure}[H]
\centering
\includegraphics[width=0.8\textwidth]{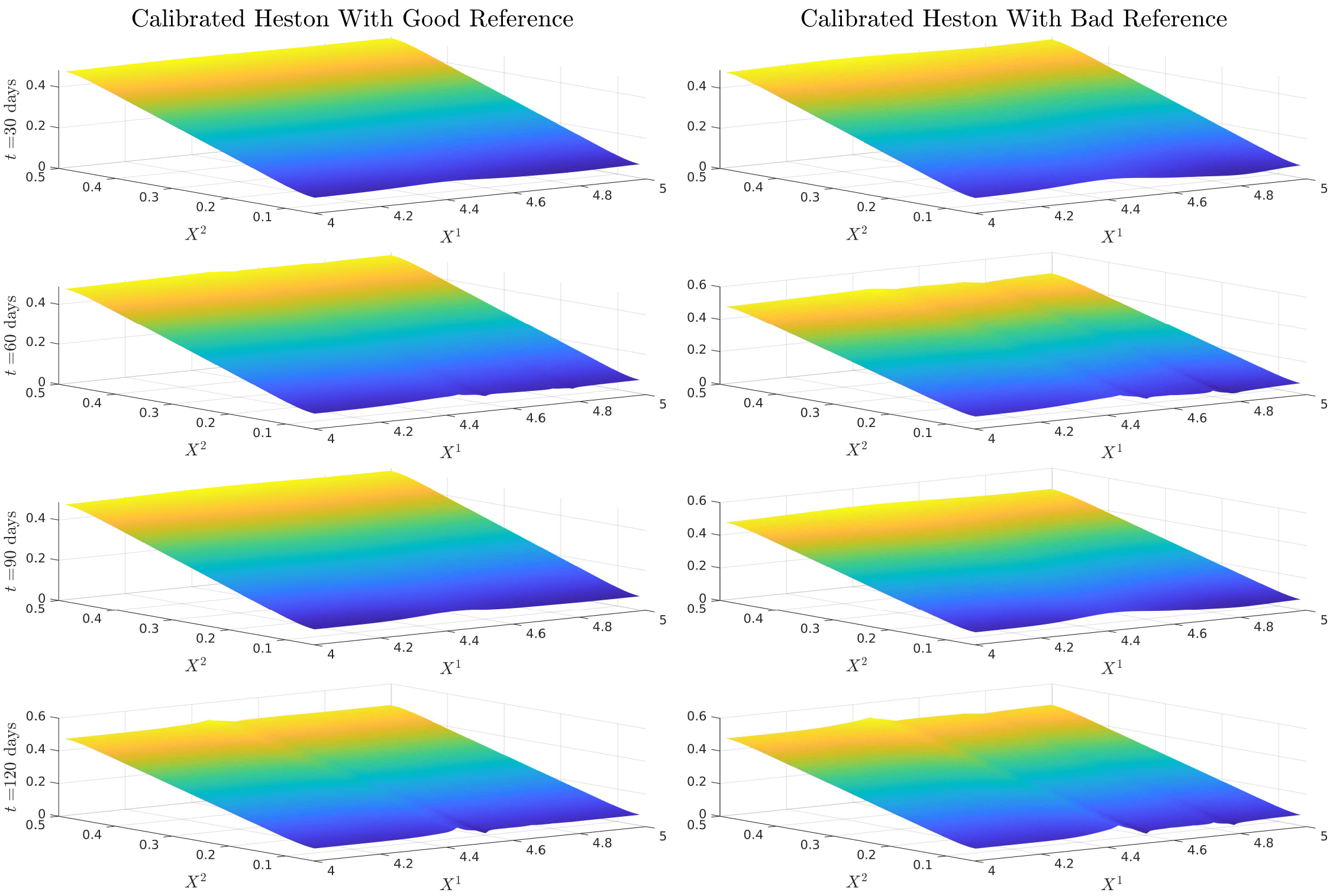} 
\caption{Plots of SOT calibrated $\beta_{11}$ under good and bad reference models. Note that the generating and reference values for $\beta_{11}$ are given by $X^2$.}
\end{figure}
\begin{figure}[H]
\centering
	\includegraphics[width=0.8\textwidth]{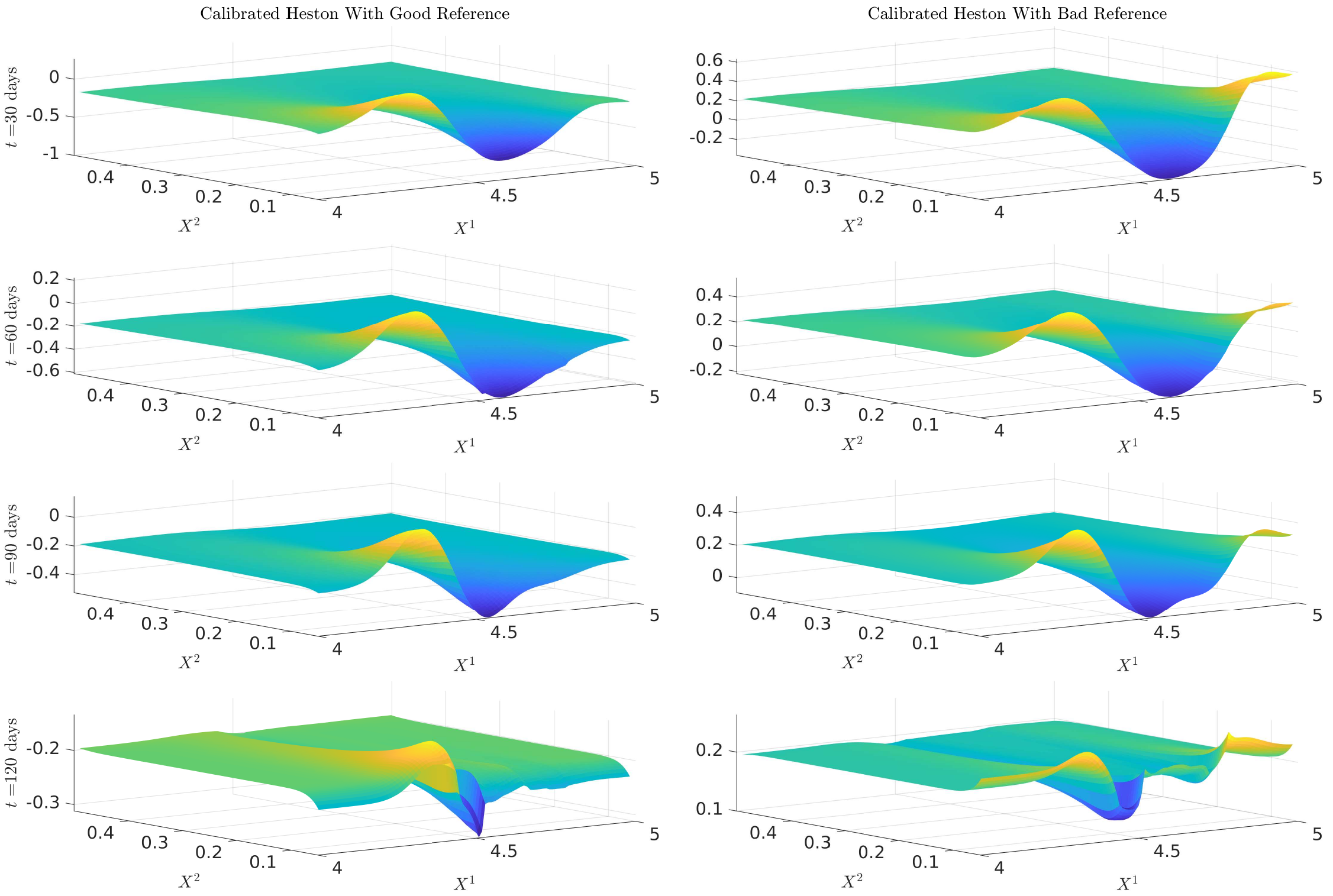} 
\caption{Plots of SOT calibrated $\rho$ under good and bad reference models. Note that $\rho=-0.4$ in the generating model, $\overline{\rho}_{\mathrm{good}}=-0.2$ and $\overline{\rho}_{\mathrm{bad}}=0.2$.}
\end{figure}
\addcontentsline{toc}{section}{References}

\printbibliography

@article{guo2022calibration,
  title={Calibration of local-stochastic volatility models by optimal transport},
  author={Guo, Ivan and Loeper, Gr{\'e}goire and Wang, Shiyi},
  journal={Mathematical Finance},
  volume={32},
  number={1},
  pages={46--77},
  year={2022},
  publisher={Wiley Online Library}
}

@article{guo2022joint,
  title={Joint Modeling and Calibration of {SPX} and {VIX} by Optimal Transport},
  author={Guo, Ivan and Loeper, Gr{\'e}goire and Ob{\l}{\'o}j, Jan and Wang, Shiyi},
  journal={SIAM Journal on Financial Mathematics},
  volume={13},
  number={1},
  pages={1--31},
  year={2022},
  publisher={SIAM}
}

@book{brigo2007interest,
  title={Interest rate models-theory and practice: with smile, inflation and credit},
  author={Brigo, Damiano and Mercurio, Fabio},
  year={2007},
  publisher={Springer Science \& Business Media}
}

@article{gyongy1986mimicking,
  title={Mimicking the one-dimensional marginal distributions of processes having an It{\^o} differential},
  author={Gy{\"o}ngy, Istv{\'a}n},
  journal={Probability Theory and Related Fields},
  volume={71},
  number={4},
  pages={501--516},
  year={1986},
  publisher={Springer}
}

@article{brunick2013mimicking,
  title={Mimicking an It{\^o} process by a solution of a stochastic differential equation},
  author={Brunick, Gerard and Shreve, Steven},
  journal={Annals of Applied Probability},
  volume={23},
  number={4},
  pages={1584--1628},
  year={2013},
  publisher={Institute of Mathematical Statistics}
}

@article{benamou2000computational,
  title={A computational fluid mechanics solution to the Monge-Kantorovich mass transfer problem},
  author={Benamou, Jean-David and Brenier, Yann},
  journal={Numerische Mathematik},
  volume={84},
  number={3},
  pages={375--393},
  year={2000},
  publisher={Springer}
}

@article{hull1990pricing,
  title={Pricing interest-rate-derivative securities},
  author={Hull, John and White, Alan},
  journal={The Review of Financial Studies},
  volume={3},
  number={4},
  pages={573--592},
  year={1990},
  publisher={Oxford University Press}
}

@article{hull1994branching,
  title={Branching out},
  author={Hull, John and White, Alan},
  journal={Risk},
  volume={7},
  number={7},
  pages={34--37},
  year={1994}
}

@article{hull1995note,
  title={A note on the models of Hull and White for pricing options on the term structure: Response},
  author={Hull, John and White, Alan},
  journal={The Journal of Fixed Income},
  volume={5},
  number={2},
  pages={97--102},
  year={1995},
  publisher={Institutional Investor Journals Umbrella}
}

@incollection{guo2019local,
  title={Local volatility calibration by optimal transport},
  author={Guo, Ivan and Loeper, Gr{\'e}goire and Wang, Shiyi},
  booktitle={2017 MATRIX Annals},
  pages={51--64},
  year={2019},
  publisher={Springer}
}

@book {rockafellar1970convex,
    AUTHOR = {Rockafellar, R. Tyrrell},
     TITLE = {Convex analysis},
    SERIES = {Princeton Mathematical Series, No. 28},
 PUBLISHER = {Princeton University Press, Princeton, N.J.},
      YEAR = {1970}
}

@article{tan2013optimal,
  title={Optimal transportation under controlled stochastic dynamics},
  author={Tan, Xiaolu and Touzi, Nizar},
  journal={The Annals of Probability},
  volume={41},
  number={5},
  pages={3201--3240},
  year={2013},
  publisher={Institute of Mathematical Statistics}
}

@article{cox1996constant,
  title={The constant elasticity of variance option pricing model},
  author={Cox, John},
  journal={Journal of Portfolio Management},
  pages={15--17},
  year={1996},
  publisher={Pageant Media}
}

@article{cox1975notes,
  title={Notes on option pricing I: Constant elasticity of variance diffusions},
  author={Cox, John},
  journal={Unpublished note, Stanford University, Graduate School of Business},
  year={1975}
}

@article{vasicek1977equilibrium,
  title={An equilibrium characterization of the term structure},
  author={Vasicek, Oldrich},
  journal={Journal of Financial Economics},
  volume={5},
  number={2},
  pages={177--188},
  year={1977},
  publisher={Elsevier}
}

@article{heston1993closed,
  title={A closed-form solution for options with stochastic volatility with applications to bond and currency options},
  author={Heston, Steven},
  journal={The Review of Financial Studies},
  volume={6},
  number={2},
  pages={327--343},
  year={1993},
  publisher={Oxford University Press}
}

@article{ma2017unconditionally,
  title={An unconditionally monotone numerical scheme for the two-factor uncertain volatility model},
  author={Ma, K and Forsyth, PA},
  journal={IMA Journal of Numerical Analysis},
  volume={37},
  number={2},
  pages={905--944},
  year={2017},
  publisher={Oxford University Press}
}

@article{liu1989limited,
  title={On the limited memory {BFGS} method for large scale optimization},
  author={Liu, Dong and Nocedal, Jorge},
  journal={Mathematical Programming},
  volume={45},
  number={1},
  pages={503--528},
  year={1989},
  publisher={Springer}
}

@article{guo2021path,
  title={Path dependent optimal transport and model calibration on exotic derivatives},
  author={Guo, Ivan and Loeper, Gr{\'e}goire},
  journal={The Annals of Applied Probability},
  volume={31},
  number={3},
  pages={1232--1263},
  year={2021},
  publisher={Institute of Mathematical Statistics}
}

@article{avellaneda1997calibrating,
  title={Calibrating volatility surfaces via relative-entropy minimization},
  author={Avellaneda, Marco and Friedman, Craig and Holmes, Richard and Samperi, Dominick},
  journal={Applied Mathematical Finance},
  volume={4},
  number={1},
  pages={37--64},
  year={1997},
  publisher={Taylor \& Francis}
}

@article{guo2022optimal,
  title={Optimal transport for model calibration},
  author={Guo, Ivan and Loeper, Gr{\'e}goire and Ob{\l}{\'o}j, Jan and Wang, Shiyi},
  journal={Risk Magazine},
  year={2022},
  publisher={Risk. net}
}

@article{lions1983optimal,
  title={Optimal control of diffusion processes and Hamilton--Jacobi--Bellman equations part 2: viscosity solutions and uniqueness},
  author={Lions, Pierre-Louis},
  journal={Communications in Partial Differential Equations},
  volume={8},
  number={11},
  pages={1229--1276},
  year={1983},
  publisher={Taylor \& Francis}
}

@misc{JLO23calibration,
      title={Calibration of Local Volatility Models with Stochastic Interest Rates using Optimal Transport}, 
      author={Joseph, Benjamin and Loeper, Gr{\'e}goire and Ob{\l}{\'o}j, Jan},
      year={2023},
      eprint={2305.00200},
      archivePrefix={arXiv},
}

@article{corrias1996fast,
  title={Fast Legendre--Fenchel transform and applications to Hamilton--Jacobi equations and conservation laws},
  author={Corrias, Lucilla},
  journal={SIAM journal on numerical analysis},
  volume={33},
  number={4},
  pages={1534--1558},
  year={1996},
  publisher={SIAM}
}

@article{lucet1997faster,
  title={Faster than the fast Legendre transform, the linear-time Legendre transform},
  author={Lucet, Yves},
  journal={Numerical Algorithms},
  volume={16},
  pages={171--185},
  year={1997},
  publisher={Springer}
}

@article{LoeperQuiros,
  title={Interior second derivative estimates for nonlinear diffusions},
  author={Loeper, Gr{\'e}goire and Quiros, Fernando},
  journal={arXiv preprint arXiv:1812.11253},
  year={2018}
}

@article {LoMi1,
    AUTHOR = {Loeper, Gr{\'e}goire},
     TITLE = {Option pricing with linear market impact and nonlinear
              {B}lack-{S}choles equations},
   JOURNAL = {Ann. Appl. Probab.},
  FJOURNAL = {The Annals of Applied Probability},
    VOLUME = {28},
      YEAR = {2018},
    NUMBER = {5},
     PAGES = {2664--2726},
      ISSN = {1050-5164},
   MRCLASS = {91G20 (35K55 49L20 60H30 93E20)},
  MRNUMBER = {3847970},
       DOI = {10.1214/17-AAP1367},
       URL = {https://doi-org.ezproxy.lib.monash.edu.au/10.1214/17-AAP1367},
}
\vspace{1cm}

\end{document}